\documentclass[preprint,pre,superscriptaddress,showpacs]{revtex4}
\usepackage{graphicx}
\usepackage{lipsum}
\usepackage{float}
\usepackage{amsmath}
\usepackage{amssymb}
\usepackage{mathtools}

\begin{document}

\title{Distribution of shortest cycle lengths in random networks}

\author{Haggai Bonneau}
\affiliation{Racah Institute of Physics, The Hebrew University, 
Jerusalem 91904, Israel}

\author{Aviv Hassid}
\affiliation{Racah Institute of Physics, The Hebrew University, 
Jerusalem 91904, Israel}

\author{Ofer Biham}
\affiliation{Racah Institute of Physics, The Hebrew University, 
Jerusalem 91904, Israel}

\author{Reimer K\"uhn}
\affiliation{Department of Mathematics,
King's College London,
Strand, London WC2R 2LS, UK}

\author{Eytan Katzav} 
\affiliation{Racah Institute of Physics, The Hebrew University, 
Jerusalem 91904, Israel}

\begin{abstract}
We present analytical results for the distribution of shortest 
cycle lengths (DSCL) in random networks.
The approach is based on the relation between the DSCL and 
the distribution of shortest path lengths (DSPL). 
We apply this approach to configuration model
networks, for which analytical results for the DSPL 
were obtained before.
We first calculate the fraction of nodes in the network which
reside on at least one cycle.
Conditioning on being on a cycle, we provide the
DSCL over ensembles of configuration model networks
with degree distributions
which follow a Poisson distribution
(Erd{\H o}s-R\'enyi network), degenerate distribution
(random regular graph) and a power-law distribution
(scale-free network).
The mean and variance of the DSCL
are calculated.
The analytical results are found to be in very good agreement 
with the results of computer simulations.
\end{abstract}

\maketitle

\section{Introduction}

Network models provide a useful conceptual framework 
for the study of a large variety of systems and processes
in science, technology and society
\cite{Havlin2010,Newman2010,Estrada2011,Barrat2012}.
These models consist of nodes and edges, where the nodes
represent physical objects, while the edges represent the
interactions between them.
Unlike regular lattices in which all the nodes have the same coordination
number, network models are characterized by a degree distribution
$P(K=k)$, $k=0,1,2,\dots$,
with a mean degree denoted by $\langle K \rangle$.
An important distinction is between networks which exhibit
a narrow degree distribution 
(such as the Poisson distribution), 
and those which exhibit a broad degree distribution,
which is typically a power-law distribution of the form 
$P(K=k) \sim k^{-\gamma}$.
The latter networks are called scale-free networks.
They exhibit some highly connected nodes, called hubs,
which are essential for the integrity of these networks, and
play a dominant role in dynamical processes.

While pairs of adjacent nodes exhibit direct connections,
the interactions between most pairs of nodes
are mediated by intermediate nodes and edges.
A pair of nodes, $i$ and $j$, may be connected by many paths
of different lengths.
However, the distance, $\ell_{ij}$, between nodes $i$ and $j$, is given by 
the length of the shortest path between them.
The mean distance between all pairs of nodes in a network is denoted by 
$\langle L \rangle$.
A central feature of random networks is the small-world property,
namely the fact that the mean distance 
scales like 
$\langle L \rangle \sim \ln N$
where $N$ is the network size
\cite{Milgram1969,Watts1998,Chung2002,Chung2003}.
Moreover, it was shown that scale-free networks may be
ultrasmall depending on the exponent $\gamma$.
In particular, for $2 < \gamma < 3$, 
their mean distance scales like
$\langle L \rangle \sim \ln \ln N$
\cite{Cohen2003}.
 
The distribution of shortest path lengths (DSPL) 
between all pairs of nodes in a network 
is a fundamental property of the network structure.
The DSPL regulates the temporal evolution of 
dynamical processes on networks, 
such as signal propagation
\cite{Maayan2005}, 
navigation 
\cite{Dijkstra1959,Delling2009,Abraham2013}
and epidemic spreading 
\cite{Satorras2001,Satorras2015}.
Properties of the DSPL have been studied in different types of networks
\cite{Newman2001,Blondel2007,Dorogotsev2003,Hofstad2007,Hofstad2008,Esker2008,Shao2008,Shao2009}.
However, in spite of its importance it has not attracted nearly as much attention
as the degree distribution.

Recently, an analytical approach was developed for calculating 
the DSPL 
\cite{Katzav2015}
in the
Erd{\H o}s-R\'enyi (ER) network,
which is the simplest mathematical model of a random network
\cite{Erdos1959,Erdos1960,Erdos1961}. 
The study of the DSPL was later extended to other network models
\cite{Nitzan2016,Melnik2016,Steinbock2017}.
Using recursion equations, analytical results for 
the DSPL were obtained in different regimes, including sparse and dense 
networks of small as well as asymptotically large sizes.
The resulting distributions were found to 
be in good agreement with the results of computer simulations. 

ER networks are random graphs which exhibit
a Poisson degree distribution, with no degree-degree correlations
between pairs of adjacent nodes. 
In fact, ER networks can be considered as a 
maximum entropy ensemble, under the constraint
that the mean degree is fixed.
Moreover, the broader class of configuration model networks
generates maximum entropy ensembles
under conditions in which the entire degree distribution is constrained
\cite{Newman2010,Newman2001,Fronczak2004,Molloy1995,Molloy1998}.
For any given degree distribution, one can produce an ensemble 
of configuration model networks and perform a statistical analysis
of its properties.
Therefore, the configuration model provides a 
powerful platform for the analysis of random networks.
It is the ideal model to use as a null model
when one tries to analyze an empirical network of which the
degree distribution is known.
To this end, one constructs configuration model networks 
of the same size and the same degree distribution as 
the empirical network.
Properties of interest 
such as the DSPL
\cite{Giot2003},
the betweenness centrality
\cite{Goh2003}
and the abundance of network motifs
\cite{Milo2002,Caldarelli2004,Klemm2006}
are compared between the two networks. 
The discrepancies provide a rigorous test of the
systematic features of the empirical network versus the 
corresponding ensemble of random networks.

In addition to open paths between pairs of distinct nodes, 
networks may exhibit cycles, namely closed paths which return to their initial nodes.
The length of a cycle is given by the number of edges (or nodes) which reside
along the cycle. The shortest possible cycle is the triangle, of length $\ell=3$. 
The longest possible cycle is a Hamiltonian cycle of length $\ell=N$.
Some nodes in a network may not reside on any cycle.
Other nodes may reside on one or more cycles.
In the latter case, the shortest among these cycles is
of particular importance.
The shortest cycle on which a given node resides provides the shortest
feedback loop for signals originated from that node
and the strongest correlations between signals 
reaching the node via different links.
Therefore, the distribution of shortest cycle lengths (DSCL)
provides useful information on
chemical networks
\cite{Gleiss2001},
biological networks
\cite{Klamt2009},
feedback processes
\cite{Zanudo2017}, 
oscillations
\cite{Vladimirov2012,Goldental2015,Goldental2017}
and synchronization
\cite{Barrat2012}
in complex networks, as well as for
ranking of nodes
\cite{Kerrebroeck2008,Giscard2017}.
Moreover, the partition functions of statistical physics models
on networks can be expressed in terms of the combinatorial 
properties of the cycles, using high temperature expansions
and low temperature expansions
\cite{Yeomans1992}.

An important class of networks consists of tree networks, 
in which any pair of nodes is connected by a single path.
Thus, in tree networks the shortest path between any pair of
nodes is the only path between them and there are no cycles.
Tree structures appear in the dilute limit of random networks 
such as the ER network and the configuration model network,
below the percolation transition.
Above the percolation transition long cycles start to emerge
in the giant cluster. 
As the network becomes more strongly connected,
the size of the giant cluster increases and
the cycles become more numerous and shorter.

In this paper we present analytical results for the DSCL in
configuration model networks.
We first calculate the probability that a random node resides
on at least one cycle.
We then calculate the DSCL for all the nodes which reside
on at least one cycle.
We apply this approach to networks with Poisson,
degenerate and power-law degree distributions. 
It is found that the analytical results are in very good agreement
with numerical simulations.
Using the tail-sum formula we calculate the mean and the variance
of the DSCL for these networks. 

The paper is organized as follows.
In Sec. II we present the configuration model.
In Sec. III we consider the percolation transition
and the giant cluster in configuration model networks.
In Sec. IV we consider properties of the DSPL to be used in the
calculation of the DSCL.
In Sec. V we present analytical results for the fraction of nodes
which reside on at least one cycle.
In Sec. VI we present analytical results for the DSCL
of configuration model networks, expressed in terms of the degree distributions
and the DSPL.
In Sec. VII we apply these results to ER networks, regular graphs and 
scale-free networks.
The results are discussed in Sec. VIII and summarized in Sec. IX.
In Appendix A we present the short-distance behavior of 
the DSPL between pairs of nodes of given degrees.
In Appendices B, D and E we summarize the properties of the giant
clusters in ER networks, random regular graphs and scale-free networks,
respectively.  In Appendix C we provide some explicit expressions for 
the probabilities that random nodes of given degrees reside on at least one cycle.

\section{The configuration model}

The configuration model is a maximum entropy ensemble of
networks under the condition that the degree distribution
is imposed
\cite{Newman2001,Newman2010}.
Here we focus on the case of undirected networks, in 
which all the edges are bidirectional.
To construct such a network of $N$ nodes, one can draw
the degrees of all nodes from a desired degree
distribution, 
$P(K=k)$,
producing a degree sequence of the form
$\{ k_i \}_{i=1,\dots,N}$
(where $\sum k_i$ must be even).
The mean degree over the ensemble of networks is
$\langle K \rangle=\sum_k k P(K=k)$.
For brevity, in the rest of the paper
we use a more compact notation, 
in which $P(K = k)$
is replaced by $P(k)$, 
except for a few places in which the more
detailed notation is needed for clarity.

A convenient way to construct a configuration model network 
is to prepare the $N$ nodes such that each node, $i$, is 
connected to $k_i$ half edges
\cite{Newman2010}.
Pairs of half edges 
from different nodes
are then chosen randomly
and are connected to each other in order
to form the network. 
The result is a network with the desired degree sequence but
no correlations.
Note that towards the end of the construction
the process may get stuck.
This may happen in the case in which the only remaining pairs of half edges
belong to the same node or to pairs of nodes which are already connected to each other.
In such cases one may perform some random reconnections 
in order to enable completion of the construction.

\section{The percolation transition and the giant cluster}

Configuration model networks generically consist of many connected 
components. In some cases the size of the largest component
scales linearly with the network size, $N$. In such cases, the largest
component is called a giant cluster.
All the other components are non-extensive and are called 
finite or isolated components, and below are referred to as 
non-giant components. 
The size of the giant cluster is determined by the 
degree distribution, $P(k)$.
Some families of degree distributions can be parametrized such
that in a certain range of parameters there is no giant cluster,
while in the complementary range there is a giant cluster.
On the boundary between these two domains in the parameter
space there is a 
phase transition, which is referred to as a
percolation transition.

Consider a configuration model network 
of $N$ nodes with a given degree distribution $P(k)$.
In this paper we will employ two different sampling
procedures. The degrees of nodes which are sampled randomly 
from the network follow the overall degree distribution $P(k)$.
However, nodes which are sampled as random neighbors 
of random nodes follow a modified degree distribution,
which takes the form

\begin{equation}
{\widetilde P}(k) = \frac{k}{\langle K \rangle} P(k).
\label{eq:tilde}
\end{equation}

\noindent
This is due to the fact that such nodes are selected proportionally
to their degrees.
Each one of these degree distributions has a generating function
associated with it.
The generating function of $P(k)$ is

\begin{equation}
G_0(x) = \sum_{k=0}^{\infty}  P(k) x^{k},
\label{eq:G0}
\end{equation}

\noindent
while the generating function of $\widetilde P(k)$ is

\begin{equation}
G_1(x) = \sum_{k=0}^{\infty} {\widetilde P}(k) x^{k-1}.
\label{eq:G1}
\end{equation}

\noindent
From the definitions of 
$G_0(x)$ and $G_1(x)$ in Eqs.
(\ref{eq:G0})
and 
(\ref{eq:G1}),
respectively,
we find that
$G_0(1)=1$
and
$G_1(1)=1$.
In some networks there are no isolated nodes (of degree $k=0$)
and no leaf nodes (of degree $k=1$). 
In such networks
$P(k) > 0$ only for $k \ge 2$. 
For these networks
we find that
$G_0(0) = 0$ 
and
$G_1(0)=0$.
This implies that in such networks both $x=0$ and $x=1$ are fixed points
of both $G_0(x)$ and $G_1(x)$.

In what follows we review the well known analysis of the percolation probability 
in configuration model networks, following Refs.  
\cite{Havlin2010,Newman2010}.
Our main motivation for 
doing so is that it allows us to highlight two lesser known facts about the problem, 
which we will need in our evaluation of the DSCL below. These concern the 
degree-dependent probabilities of randomly chosen nodes and randomly 
chosen neighbors of randomly chosen nodes to belong to the giant cluster. 
The probability that a random node resides on the giant
cluster is denoted by $g$. In the case in which a giant cluster
exists, $g > 0$, while in the case in which there is no giant cluster, $g=0$. 
To obtain the probability $g$, 
one needs to first calculate the probability $\tilde g$
that a random neighbor of a random node, $i$, 
belongs to the giant cluster in the reduced network, 
which does not include the node $i$.
In the thermodynamic limit, $N \rightarrow \infty$,
the probability $\tilde g$ is given as a solution of the
self-consistency equation
\cite{Havlin2010}

\begin{equation}
1 - {\tilde g} = G_1(1 - {\tilde g}).
\label{eq:tg}
\end{equation}

\noindent
The left hand side of this equation is the probability that a random neighbor of
a random node does not reside on the giant cluster.
The right hand side represents the same quantity in terms of its neighbors, namely as the probability that
none of the neighbors of such node resides on the giant cluster.
Once $\tilde g$ is known, the probability $g$ can be obtained from

\begin{equation}
g = 1 - G_0(1 - {\tilde g}).
\label{eq:g}
\end{equation}

\noindent
This relation is based on the same consideration as Eq. (\ref{eq:tg}),
where the difference is that the reference node is a random node rather
than a random neighbor of a random node.

Below we consider the more specific case of nodes of a given degree.
The probability that a random node of a given degree, $k$, resides on the 
giant cluster is denoted by $g_k$.
Using the degree distribution, $P(k)$, the
probability, $g$, that a random 
node of an unspecified degree resides on the giant cluster can
be expressed in terms of $g_k$ by

\begin{equation}
g = \sum_{k=0}^{\infty} P(k) g_k.
\label{eq:g_k}
\end{equation}

\noindent
Such a node resides on the giant cluster if at least one of its $k$ neighbors
resides on the giant cluster. 
Therefore,

\begin{equation}
g_k = 1 - (1 - {\tilde g})^k.
\label{eq:g_kp}
\end{equation}

\noindent
Thus, high degree nodes are more likely to reside on the giant cluster than
low degree nodes.
Similarly, the probability $\tilde g$ that a random neighbor of a random node resides
on the giant cluster can be expressed in the form

\begin{equation}
{\tilde g} = \sum_{k=0}^{\infty} \widetilde P(k) {\tilde g}_k,
\end{equation}

\noindent
where $\tilde g_k$ is the probability that
a random neighbor of a random node resides on the giant cluster,
under the condition that its degree is $k$.
Using similar considerations, 
we find that the probability $\tilde g_k$ is given by

\begin{equation}
\tilde g_k = 1 - (1 - \tilde g)^{k-1}.
\label{eq:tg_k}
\end{equation}

\noindent
In Appendices B, D and E we apply these considerations to the analysis of
the giant clusters in ER networks, random regular graphs and scale-free
networks, respectively. 

\section{The distribution of shortest path lengths}

Consider a pair of random nodes, $i$ and $j$, in a 
network of $N$ nodes. Assuming that the two nodes
reside on the same connected component, they may be
connected to each other by a large number of paths. 
The distance between the two nodes is equal to
the length of the shortest among these paths 
(possibly more than one). 
Below we briefly review the approach introduced in Ref.
\cite{Nitzan2016} 
for the calculation of the DSPL
in configuration model networks of a given size, $N$, and a given degree distribution,
$P(k)$.
The DSPL can be expressed in the form of a tail distribution,
where $P_{\rm PL}(L>\ell)$ is the probability
that the shortest path length between a random pair of nodes
is larger than $\ell$.
The tail distribution can be expressed as a product of the form

\begin{equation}
P_{\rm PL}(L>\ell) = 
\prod_{\ell^{\prime}=1}^{\ell}
P_{\rm PL}(L>\ell^{\prime} | L>\ell^{\prime}-1),
\label{eq:ProdCond}
\end{equation}

\noindent
where $P_{\rm PL}(L>\ell | L>\ell-1)$
is the conditional probability that the distance between
a random pair of nodes is larger than $\ell$ conditioned on
it being larger than $\ell-1$.
In the analysis below we use different types of tail distributions
for the DSPL. In Table I we summarize these distributions and
list the equations from which each one of them can be evaluated.

A path 
of length $\ell$
from node $i$ to node $j$ can be decomposed 
into a single edge connecting node $i$ and node 
$r \in \partial_i$
(where $\partial_i$ is the set of all nodes directly connected to $i$),
and a shorter path of length 
$\ell-1$ connecting $r$ and $j$.
Thus, the existence of a path of length $\ell$
between nodes $i$ and $j$
can be ruled out if there is no path of length
$\ell-1$ between any of the nodes 
$r \in \partial_i$,
and $j$.
For sufficiently large networks,
the argument presented above translates into
the recursion equation
\cite{Nitzan2016}

\begin{equation}
P_{\rm PL}(L > \ell | L>\ell-1) =
G_0[{\widetilde P}_{\rm PL}(L>\ell-1 | L> \ell-2)],
\label{eq:P_rec2}
\end{equation}

\noindent
where the generating function $G_0(x)$ is given by
Eq. (\ref{eq:G0}).
Here we distinguish between the conditional probability
$P_{\rm PL}(L>\ell | L>\ell-1)$ between nodes $i$ and $j$
and the probability
${\widetilde P}_{\rm PL}(L>\ell | L>\ell-1)$ 
between a node $r \in \partial_i$
and node $j$, on the reduced network from which node $i$ was removed.
The reason for this distinction is that the former probability involves two 
random nodes, while the latter probability involves a node, $r$, which is 
a random neighbor of a random node, and a random node, $j$.
The conditional probability 
${\widetilde P}_{\rm PL}(L>\ell | L>\ell-1)$ 
satisfies the recursion equation

\begin{equation}
{\widetilde P}_{\rm PL}(L>\ell | L>\ell-1) =
G_1[ {\widetilde P}_{\rm PL}(L>\ell-1 | L> \ell-2)],
\label{eq:P_rec2s}
\end{equation}

\noindent
where $G_1(x)$ is given by 
Eq. (\ref{eq:G1}),
which is valid for 
$\ell \ge 2$.

The case of $\ell=1$ deserves special attention.
On a network of size $N$ (sufficiently large),
the probability that two random nodes 
are not connected is given by
\cite{Nitzan2016}

\begin{equation}
P_{\rm PL}(L>1 | L>0) \simeq 1 - \frac{\langle K \rangle}{N-1} 
+ {\cal O} \left(\frac{1}{N^2}  \right),
\label{eq:m1}
\end{equation}

\noindent
while
the probability that a random neighbor of a random node and a random node
are not connected is given by

\begin{equation}
{\widetilde P}_{\rm PL}(L>1 | L>0) \simeq 1 - 
\frac{\langle K^2 \rangle - \langle K \rangle}{\langle K \rangle (N-1)}
+ {\cal O} \left(\frac{1}{N^2}  \right).
\label{eq:tilde_m1}
\end{equation}

\noindent
The difference between Eqs.
(\ref{eq:m1})
and
(\ref{eq:tilde_m1})
is due to the fact that the degree distribution 
${\widetilde P}(k)$ of 
random neighbors of random nodes,
given by Eq. (\ref{eq:tilde}),
is generically distinct from the degree
distribution 
$P(k)$
of random nodes.

Actually, there are two other types of DSPLs in random networks,
which are needed for the analysis of shortest cycles. One of them is the
DSPL between a random node and a random neighbor of a random node,
denoted by ${\widetilde P}_{\rm PL}(L>\ell)$. 
The other one is the DSPL between
two random neighbors of random nodes,
denoted by $\widehat P_{\rm PL}(L>\ell)$.
The DSPL between a random node and a 
random neighbor of a random node is expressed
as a product of the form

\begin{equation}
\widetilde P_{\rm PL}(L>\ell) = \prod_{\ell^{\prime}=1}^{\ell}
\widetilde P_{\rm PL}(L>\ell^{\prime} | L>\ell^{\prime}-1),
\label{eq:prodtilde}
\end{equation}

\noindent
where
$\widetilde P_{\rm  PL}(L>\ell^{\prime}| L>\ell^{\prime}-1)$
is obtained by iterating Eq.
(\ref{eq:P_rec2s}),
using Eq. (\ref{eq:tilde_m1})
as an initial condition.

The DSPL between two 
random neighbors of random nodes,
$\widehat P_{\rm PL}(L>\ell)$,
requires a careful attention.
The initial condition in this case, namely the probability 
that two such nodes are not connected 
on a network of size $N$ is

\begin{equation}
\widehat P_{\rm PL}(L>1 | L>0) 
=
\sum_{k=0}^{\infty} 
%\frac{k}{\langle K \rangle} P(k) 
\widetilde P(k)
\sum_{k'=0}^{\infty}
%\frac{k'}{\langle K \rangle} P(k')
\widetilde P(k')
\left[ 1 - \frac{k'-1}{(N-1) \langle K \rangle} \right]^{k-1}. 
\label{eq:hat_m1}
\end{equation}

\noindent
Using a binomial approximation 
and performing the summations,
we obtain

\begin{equation}
\widehat P_{\rm PL}(L>1 | L>0) = 
1 -
\frac{\langle K \rangle}{N-1} 
\left( \frac{ \langle K^2 \rangle - \langle K \rangle}{\langle K \rangle^2} \right)^2
+ {\cal O} \left(\frac{1}{N^2}  \right).
\label{eq:L1hat}
\end{equation}

\noindent
This initial condition is fed into the recursion equation

\begin{equation}
\widehat P_{\rm PL}(L>\ell | L>\ell-1) =
G_1[ \widehat P_{\rm PL}(L>\ell-1 | L> \ell-2)].
\label{eq:P_rec2sh}
\end{equation}

\noindent
The DSPL between random neighbors of random nodes is then obtained
as a product of the conditional probabilities:

\begin{equation}
\widehat P_{\rm PL}(L>\ell) = \prod_{\ell^{\prime}=1}^{\ell}
\widehat P_{\rm PL}(L>\ell^{\prime} | L>\ell^{\prime}-1).
\label{eq:prodh}
\end{equation}

In the analysis above, we considered only pairs of nodes which
reside on the same cluster.
Since not all pairs of random nodes reside on the same cluster,
the DSPL needs to be adjusted.
Taking a random pair of nodes, $i$ and $j$, the probability
that they reside on the same cluster
is negligible, unless they both reside on the giant cluster.
The probability that both nodes reside on the giant cluster is $g^2$.
Therefore, the probability that the distance between them is
infinite is $P_{\rm PL}(L=\infty) = 1-g^2$.
This implies that the DSPL between
all pairs of nodes in the network (without assuming that they
reside on the same cluster) is

\begin{equation}
Q_{\rm PL}(L>\ell) = g^2 P_{\rm PL}(L>\ell) + (1-g^2).
\label{eq:Q}
\end{equation}

\noindent
Using a similar argument for the DSPL between a random node
and a random neighbor of a random node, we find that the
DSPL between all such pairs is given by

\begin{equation}
\widetilde Q_{\rm PL}(L>\ell) = g {\tilde g} \widetilde P_{\rm PL}(L>\ell) + (1 - g \tilde g).
\label{eq:Qt}
\end{equation}

\noindent
Similarly, the DSPL between all pairs of 
random neighbors of random nodes is given by

\begin{equation}
\widehat Q_{\rm PL}(L>\ell) =  {\tilde g}^2 \widehat P_{\rm PL}(L>\ell) + (1 -  {\tilde g}^2).
\label{eq:Qw}
\end{equation}

\noindent
In cases where $g<1$, the overall DSPLs,
$Q(L>\ell)$, $\widetilde Q(L>\ell)$ and $\widehat Q(L>\ell)$,
approach a non-zero asymptotic
value at large $\ell$, unlike the original DSPLs, 
$P(L>\ell)$, $\widetilde P(L>\ell)$ and $\widehat P(L>\ell)$,
which decay to zero.

To obtain the DSPL between random pairs of nodes of known degrees,
consider two random nodes, $i$ and $j$, of degrees $k$ and $k'$, respectively,
which do not share any neighbors and thus the distance between them
satisfies $\ell > 2$.
Since node $i$ has $k$ neighbors and node $j$ has $k'$ neighbors, the probability
that the distance between them is longer than $\ell$
is equal to the probability that
the distance between any neighbor of $i$ to 
any neighbor of $j$ is longer than
$\ell-2$. 
Therefore,

\begin{equation}
P_{\rm PL}(L > \ell | k,k') = [ \widehat P_{\rm PL}(L>\ell-2)]^{k k'},
\label{eq:P}
\end{equation}

\noindent
where $\widehat P_{\rm PL}(L>\ell)$ is 
the DSPL between two random neighbors of random nodes,
given by
Eq. (\ref{eq:prodh}).
Similarly, the DSPL between a random node, of degree $k$, and a random 
neighbor of a random node,
of degree $k'$, is given by

\begin{equation}
\widetilde P_{\rm PL}(L > \ell |  k,k') = [ \widehat P_{\rm PL}(L>\ell-2)]^{k(k'-1)}.
\label{eq:wtP}
\end{equation}

\noindent
The DSPL between pairs of random neighbors of random nodes, 
under the condition that their degrees are $k$ and $k'$,
is given by

\begin{equation}
\widehat P_{\rm PL}(L > \ell | k,k') 
= 
[ \widehat P_{\rm PL}(L>\ell-2)]^{(k-1) (k'-1)}.
\label{eq:whP}
\end{equation}

\noindent
It is important to note that
Eqs. (\ref{eq:P})-(\ref{eq:whP})
are valid for $\ell>2$.
The corresponding equations for the conditional
probabilities 
$P_{\rm PL}(L > \ell | k,k')$,
$\widetilde P_{\rm PL}(L > \ell | k,k')$
and
$\widehat P_{\rm PL}(L > \ell | k,k')$,
with $\ell=1,2$
are presented in Appendix A.

Using the results presented above we now provide the overall DSPLs, 
between random pairs of nodes of
known degrees. Considering two random nodes 
of degrees $k$ and $k'$, we obtain

\begin{equation}
Q_{\rm PL}(L>\ell | k,k') = g_{k} g_{k'} 
P_{\rm PL}(L>\ell | k,k') 
+ (1 - g_{k} g_{k'}),
\label{eq:Q2}
\end{equation}

\noindent
where $g_k$ is  given by 
Eq. (\ref{eq:g_kp}).
Similarly, the DSPL between a random node 
of degree $k$, and a random neighbor of a random 
node, of degree $k'$ is given by

\begin{equation}
\widetilde Q_{\rm PL}(L>\ell | k,k') = g_k {\tilde g}_{k'} 
\widetilde P_{\rm PL}(L>\ell | k, k') + (1 - g_{k} {\tilde g}_{k'}),
\label{eq:Qtc}
\end{equation}

\noindent
where $\tilde g_{k'}$ is 
given by Eq. (\ref{eq:tg_k}).
Lastly, the DSPL between pairs of random neighbors of random nodes,
conditioned on their degrees, $k$ and $k'$,
is given by

\begin{equation}
\widehat Q_{\rm PL}(L>\ell | k, k') = {\tilde g}_{k} {\tilde g}_{k'} 
\widehat P_{\rm PL}(L>\ell | k, k') + (1 - \tilde g_{k} \tilde g_{k'}).
\label{eq:Qhc}
\end{equation}

\noindent
The moments of
$P_{\rm PL}(L>\ell)$
provide useful information about the network.
The $n^{\rm th}$ moment, 
$\langle L^n \rangle_{\rm PL}$,
can be obtained 
using the tail-sum formula
\cite{Pitman1993}

\begin{equation}
\langle L^n \rangle_{\rm PL} = 
\sum_{\ell=0}^{N-2} [(\ell+1)^n - \ell^n] P_{\rm PL}(L>\ell).
\label{eq:tail_sum}
\end{equation}

\noindent
Note that the sum in 
Eq.
(\ref{eq:tail_sum})
does not extend to $\infty$ because 
the longest possible shortest path in a 
network of size $N$ is $N-1$.
The mean distance in configuration model
networks has been studied extensively
\cite{Newman2001,Chung2002,Chung2003,Hofstad2007,Esker2008}.
It was found that 

\begin{equation}
\langle L \rangle_{\rm PL}
\simeq 
\frac{\ln N}{\ln \left(\frac{\langle K^2 \rangle - \langle K \rangle}
{\langle K \rangle}\right)} + {\cal O}(1). 
\label{eq:mean_ellold}
\end{equation}

\noindent
The width of the distribution 
can be characterized by the
variance
$\sigma_{\rm PL}^2 
= 
\langle L^2 \rangle_{\rm PL} 
- 
\langle L \rangle_{\rm PL}^2$.

\section{The fraction of nodes which reside on at least one cycle}

In this section we calculate the probability 
$P(i \in {\rm cycle})$
that a random node, $i$,
resides on at least one cycle.
To do so, we first calculate the conditional probability,
$P(i \in {\rm cycle} | k)$,
that a node of a given degree, $k$,
resides on at least one cycle.
Actually, this probability can be expressed by
$P(i \in {\rm cycle} | k ) = 1 - P(i \notin {\rm cycle} |k)$.
Clearly, nodes of degree $k=0$ or $1$ cannot reside on
any cycle and thus

\begin{equation}
P(i \notin {\rm cycle} | 0)=P(i \notin {\rm cycle} | 1) =1.
\end{equation}

\noindent 
For a node of degree $k \ge 2$ to reside on a cycle,
two of its neighbors must be connected to each other on 
the reduced network from which $i$ is removed.
The probability that a neighbor of a random node $i$, on the
reduced network from which $i$ is removed, is part of the
giant cluster of the reduced network is equal to $\tilde g$.
The probability that a given pair of neighbors will reside on
the giant cluster of the reduced network is ${\tilde g}^2$.
This pair of neighbors will reside on the same component only
if this component is the giant cluster (up to negligible probability).
Hence, the probability that a given pair of neighbors of $i$ is
not connected is $1 - {\tilde g}^2$.
Since there are $\binom{k}{2}$ pairs of neighbors of node $i$,
the probability that none of these pairs are connected on the
reduced network from which node $i$ is removed is

\begin{equation}
P(i \notin {\rm cycle} |k) = \left( 1 - {\tilde g}^2 \right)^{\binom{k}{2}}.
\label{eq:cyck}
\end{equation}

\noindent
Note that this result is based on the assumption that the paths 
between different pairs of neighbors of $i$ are independent.
This assumption is expected to hold in ensembles of uncorrelated networks,
such as the configuration model or any other network model in which
the clustering coefficient is small.

Using the arguments discussed above
we find that
the probability that a random node of unspecified degree
resides on at least one cycle is given by

\begin{equation}
P(i \in {\rm cycle})
=
1
- \left[P(K=0)+P(K=1)
+\sum_{k=2}^{\infty} P(k)
P(i \notin {\rm cycle} | k)
\right],
\label{eq:cyc}
\end{equation}

\noindent
where $P(i \notin {\rm cycle} | k)$
is given by Eq. (\ref{eq:cyck}).
Note that in the case of $\tilde g=0$ one can show,
using Eq. (\ref{eq:g}) that $g=0$ as well, meaning
that there is no giant cluster.
Eq. (\ref{eq:cyc}) shows that under these conditions
$P(i \in {\rm cycle})=0$.
This reflects the fact that for a network below the percolation threshold,
in the thermodynamic limit, the number of cycles does not scale with
$N$ \cite{Takacs1988,Bollobas2001}.
Thus, essentially all the components are trees.

One should point out that Eq. (\ref{eq:cyck}) does not take into account
certain correlations between pairs of neighbors of node $i$.
To demonstrate this point,
consider the $k$ neighbors of node $i$. We will denote their degrees by
$k_1,k_2,\dots,k_k$. These degrees are independent of each other and
are all drawn from the same distribution, $\widetilde P(k)$.
However, the probability that a node, $r_m$, resides on the giant
cluster depends on its degree, $k_m$,
and is given by $g_{k_m}$
[Eq. (\ref{eq:g_kp})].
Since each neighbor of $i$ participates in $k-1$ such pairs, the
probabilities that different pairs reside on the giant cluster are
not independent.
Each one of these nodes,  may connect to each of the other $k-1$
neighbors, with a probability which depends on its degree. 
Therefore, these $k-1$ probabilities are not independent, unlike
the assumption made in Eq. (\ref{eq:cyck}).
To account for these correlations we  express $P(i \notin {\rm cycle} |k)$
in the form

\begin{equation}
P(i \notin {\rm cycle} | k) = 
\sum_{k_1,k_2,\dots,k_k}
\prod_{r=1}^k
%\frac{k_r}{\langle K \rangle} P(k_r)
\widetilde P(k_r)
\prod_{m < n} 
(1 - {\tilde g}_{k_m} {\tilde g}_{k_n})
\label{eq:Pncc2}
\end{equation}

\noindent
where 
the product runs over all pairs of neighbors of node $i$.

In summary, we have presented two approaches to the calculation
of $P(i \notin {\rm cycle})$.
The simpler approach of Eq. (\ref{eq:cyck})
provides a good approximation in most cases.
For highly heterogeneous networks one may need the more
detailed approach of Eq. (\ref{eq:Pncc2}),
which is much more elaborate to implement.
More specifically, it requires summation over all possible
degree sequences of length $k$, which becomes prohibitive
when $k$ is large.

\section{The distribution of shortest cycle lengths}

\begin{figure}
\includegraphics[width=9cm]{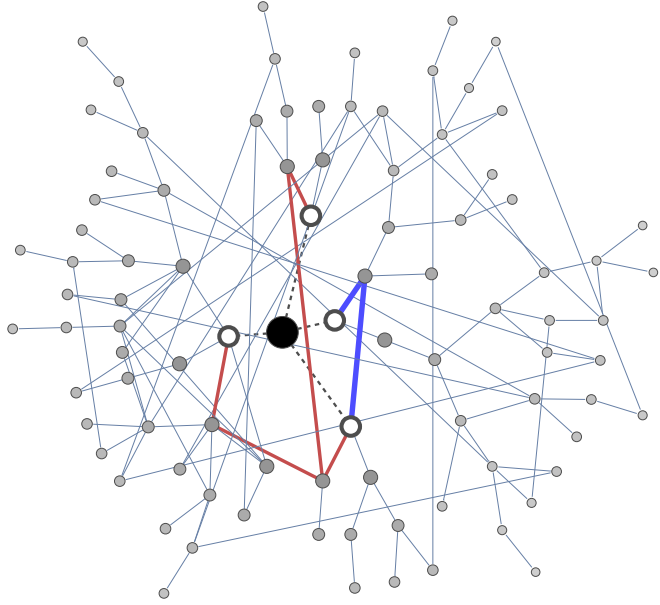}
\caption{
(Color online)
Illustration of the cycles on
which a random reference node (black filled circle) resides. The reference node
has $4$ neighbors (empty circles). 
The edges connecting the reference
node to its neighbors are shown by dashed lines. 
The paths connecting
pairs of neighbors are shown by solid lines.
The shortest path, shown by a thick solid line (blue) is of length $2$, 
thus the shortest cycle on which the reference node resides is of length $\ell=4$. 
The other paths between neighbors of the reference node, 
which are of lengths $3$ and $4$ are shown by narrower solid lines (red).
They form cycles of lengths $5$ and $6$, respectively.
}
\label{fig:1}
\end{figure}

Consider a random node, $i$, in a configuration model 
network of size $N$ with degree 
distribution $P(k)$. 
A node of degree $K \ge 2$ may reside on one or more cycles.
Here we focus on the shortest among these cycles.
More specifically, we calculate the distribution of lengths of
the shortest cycles on which a random node of degree $k$ resides.
We denote the neighbors of node $i$ by $r_1,r_2,\dots,r_k$.
A cycle of length $\ell$ on which $i$ resides, consists of the edges connecting
$i$ to two of its neighbors, $r_m$ and $r_n$ and a path of length $\ell-2$
connecting $r_m$ and $r_n$. The number of possible shortest 
cycles is $\binom{k}{2}$,
namely the number of pairs of neighbors of $i$.
In Fig. \ref{fig:1} we present an illustration of the cycles on
which a random reference node (black filled circle) resides. This node
has $k=4$ neighbors (empty circles). The edges between the reference
node and its neighbors are shown by dashed lines. The paths connecting
pairs of neighbors are shown by solid lines. The shortest among thse paths
is shown by a thick solid line (blue) of length $2$, thus the
shortest cycle on which the reference node resides is of length $\ell=4$. The other paths,
of lengths $3$ and $4$ are shown by thin solid lines (red).

The tail distribution of the lengths of shortest cycles on which random nodes of degree
$k$ reside is denoted by $P_{\rm CL}(L>\ell |K= k)$.
In order that the shortest cycle will be longer than $\ell$, the distances
between all pairs of neighbors must satisfy $L > \ell-2$.
Therefore

\begin{equation}
P_{\rm CL}(L>\ell | k) = {\widehat Q}_{\rm PL}(L>\ell-2)^{\binom{k}{2}},
\label{eq:dscl_c1}
\end{equation}

\noindent
where $\widehat Q_{\rm PL}(L>\ell)$ is given by
Eq. (\ref{eq:Qw}).
This equation is based on the assumption that the distances between
all pairs of neighbors of node $i$ are independent of each other.
This assumption is expected to be satisfied in configuration model networks.
Note that nodes of degrees $k=0$ and $1$ do not reside on any cycle,
and thus
$P_{\rm CL}(L>\ell |K = 0 ) = P_{\rm CL}(L>\ell |K = 1 ) =1$ 
for any
value of $\ell$.

For a random node, $i$, of unknown degree, the DSCL
is obtained by averaging over all possible degrees according to

\begin{equation}
P_{\rm CL}(L>\ell) = 
\sum_{k=0}^{\infty} P(k) P_{\rm CL}(L>\ell | k). 
\label{eq:dscl}
\end{equation}

\noindent
Writing this equation in a more explicit form, we obtain

\begin{equation}
P_{\rm CL}(L>\ell) 
= 
P(K=0) + P(K=1) 
+ 
\sum_{k=2}^{\infty} P(k) {\widehat Q}_{\rm PL}(L>\ell-2)^{\binom{k}{2}}.
\end{equation}

\noindent
This equation is expected to provide an accurate description
of the DSCL of configuration model networks,
when the degree distribution is not too broad.
The corresponding probability distribution function, 
$P_{\rm CL}(L=\ell)$
can be easily obtained by

\begin{equation}
P_{\rm CL}(L=\ell) = P_{\rm CL}(L>\ell-1) - P_{\rm CL}(L>\ell).
\end{equation}

Similarly to the discussion of 
$P(i \in {\rm cycle})$ in the previous section,
to obtain more accurate results for 
$P_{\rm CL}(L>\ell)$ in a
network which exhibits a broad degree distribution, one
needs to take into account the heterogeneity 
of the network. 
Consider the first shell around the random node, $i$,
which consists of the nodes $r_1,r_2,\dots,r_k$,
of degrees $k_1,k_2,\dots,k_k$.
The distribution of shortest path lengths between a pair of neighbors,
$r_m$ and $r_n$ depends on their degrees, $k_m$ and $k_n$.
Therefore, in this analysis one should use the conditional probabilities
$\widehat Q_{\rm PL}(L>\ell-2 | k_m,k_n)$.
The shortest cycle on which $i$ resides, consists of the shortest
path among all the paths connecting the $\binom{k}{2}$ pairs
of neighbors of $i$. Since each neighbor, such as $r_m$,
of degree $k_m$, participates in $k-1$ such pairs, these
conditional distributions are not independent.
Thus, one should properly  condition
on the degrees of pairs of neighbors.
Implementing these considerations, one can replace 
Eq. (\ref{eq:dscl_c1}) by

\begin{equation}
P_{\rm CL}(L>\ell | k) = 
\sum_{k_1,k_2,\dots,k_k}
\prod_{r=1}^k
%\frac{k_r}{\langle K \rangle} P(k_r)
\widetilde P(k_r)
\prod_{m < n} 
\widehat Q_{\rm PL}(L>\ell-2 | k_m,k_n)
\label{eq:dscl_c2}
\end{equation}

\noindent
where 
$\widehat Q_{\rm PL}(L>\ell-2 | k_m,k_n)$
is given by Eq. (\ref{eq:Qhc}).
Actually, for $\ell \ge N$ this equation coincides with 
Eq. (\ref{eq:Pncc2}).
This is due to the fact that the maximal length of a cycle is $\ell=N$.
Hence, the probability that the length of the shortest cycle is larger than $N$ is equivalent to
the probability that there is no cycle.
Plugging Eq. (\ref{eq:dscl_c2}) into Eq. (\ref{eq:dscl}),
we obtain a more accurate expression for the DSCL.

\begin{figure}
\includegraphics[width=4cm]{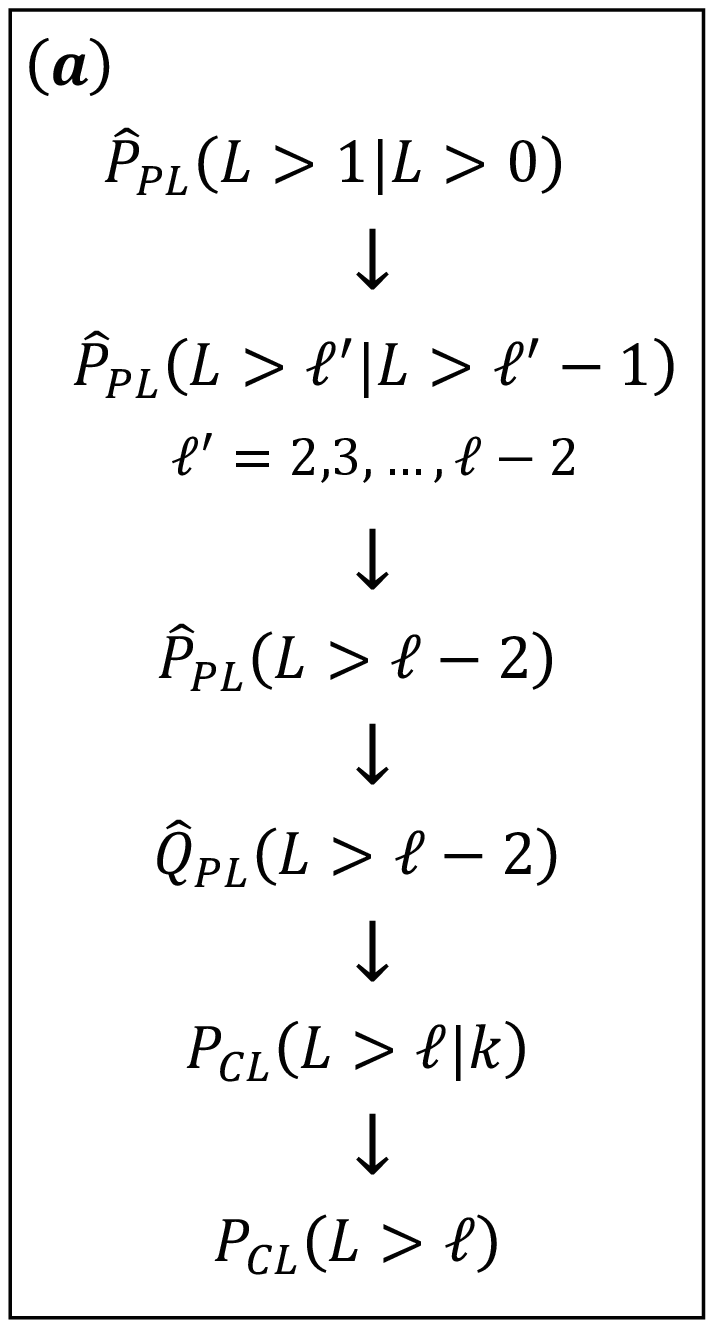}
\includegraphics[width=4cm]{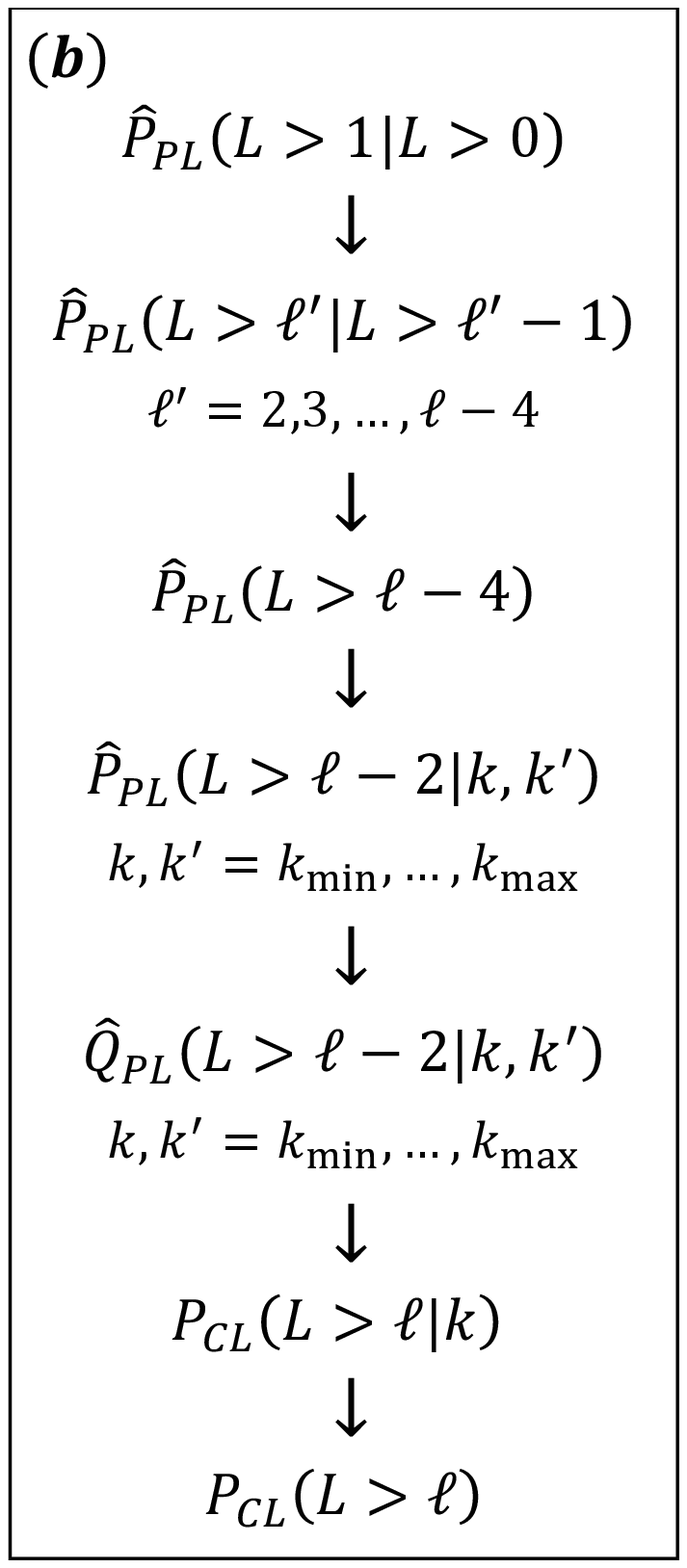}
\caption{
Flow charts illustrating the sequence of intermediate steps in the
calculation of the distribution of shortest cycle lengths,
$P_{\rm CL}(L>\ell)$: (a) in the
simpler approach of
Eq. (\ref{eq:dscl_c1}); 
(b) in the more detailed approach of
Eq. (\ref{eq:dscl_c2}).
}
\label{fig:2}
\end{figure}

In practice, for networks with broad degree distributions,
the summation over the whole range of values of $k$ and
$k_1,k_2,\dots,k_k$ may be impractical. In such cases,
one can evaluate Eq. (\ref{eq:dscl_c2}) using 
Monte Carlo methods
\cite{Newman1999}. 
The simplest approach is to draw the degree $k$ from the
distribution $P(k)$ and then draw the $k$ degree $k_1,k_2,\dots,k_k$
from the distribution $k P(k)/\langle K \rangle$.
One then calculates
$\widehat Q_{\rm PL}(L>\ell-2 | k_m,k_n)$
for all the $\binom{k}{2}$ combinations of degrees, $k_m$ and $k_n$,
and multiplies them to obtain one data point for
$P_{\rm CL}(L>\ell | k)$.
In Fig. \ref{fig:2} we present flow 
charts illustrating the sequence of intermediate steps in the
calculation of the DSCL.
The simpler approach of
Eq. (\ref{eq:dscl_c1}) is shown in Fig. \ref{fig:2}(a)
and the more detailed approach of
Eq. (\ref{eq:dscl_c2}) is shown in Fig. \ref{fig:2}(b).

The mean of the DSCL
is given by the first moment

\begin{equation}
\langle L \rangle_{\rm CL} = \sum_{\ell=2}^{N-1} P_{\rm CL}(L>\ell).
\label{eq:meanc_ell}
\end{equation}

\noindent
The variance of the DSCL is given by

\begin{equation}
\sigma_{\rm CL}^2 = \langle L^2 \rangle_{\rm CL} - \langle L \rangle_{\rm CL}^2,
\end{equation}

\noindent
where
 
\begin{equation}
\langle L^2 \rangle_{\rm CL} = \sum_{\ell=2}^{N-1} (2 \ell + 1) P_{\rm CL}(L>\ell).
\label{eq:secmomCL_ell}
\end{equation}

\noindent
Similarly, higher order moments can be obtained using
the tail-sum formula, as in Eq. (\ref{eq:tail_sum}).

\section{Applications to specific network models}

Here we apply the approach presented above for the calculation of the
DSCL in three examples of configuration model networks, namely 
ER networks, random regular graphs 
and scale-free networks.

\subsection{Erd{\H o}s-R\'enyi networks}

The Erd\H{o}s- R\'enyi (ER) network is the simplest kind of
a random network, 
and a special case of the configuration model,
in which only the mean degree,
$\langle K \rangle = c$, is constrained.
ER networks can be constructed by independently connecting
each pair of nodes with probability
$p = {c}/{(N-1)}$.
In the thermodynamic limit the resulting degree distribution follows a
Poisson distribution of the form

\begin{equation}
P(k) = \frac{e^{-c} c^k}{k!}.
\label{eq:poisson}
\end{equation}

\noindent
In Appendix B we briefly summarize the properties of the giant cluster of
the ER network and present a closed form expression
for $g$ as a function of $c$.
More generally, in ER
networks there is no distinction between 
the statistical properties of a random node and a random neighbor
of a random node.
As a result, 
$\tilde g = g$ and the different DSPLs are identical,
namely 
$P_{\rm PL}(L=\ell) 
= {\widetilde P}_{\rm PL}(L=\ell) = \widehat P_{\rm PL}(L=\ell)$.
Similarly, for the overall DSPLs we obtain
$Q_{\rm PL}(L=\ell) 
= {\widetilde Q}_{\rm PL}(L=\ell) = \widehat Q_{\rm PL}(L=\ell)$.
Inserting the degree distribution 
of Eq. (\ref{eq:poisson})
into the generating functions
$G_0(x)$ and $G_1(x)$ in 
Eqs. (\ref{eq:P_rec2}) and (\ref{eq:P_rec2s}),
respectively, one obtains the conditional probabilities
$P_{\rm PL}(L>\ell | L>\ell-1)$.
Inserting them into 
Eq. (\ref{eq:ProdCond}),
one obtains the tail DSPL
between pairs of nodes which reside on the same cluster,
denoted by $P_{\rm PL}(L > \ell)$.
This DSPL essentially accounts only for pairs of nodes which
both reside on the giant cluster, because for a pair of nodes on
the non-giant components it is extremely unlikely that they reside on the
same non-giant component.
In order to obtain the overall DSPL between all pairs of
nodes, one needs to adjust the results for the fraction of pairs of
nodes in which both of them reside on the giant cluster, which is
given by $g^2$. Inserting the probability $P_{\rm PL}(L > \ell)$
into Eq. (\ref{eq:Q}), one obtains the overall DSPL,
$Q_{\rm PL}(L>\ell)$.

\begin{figure}
\includegraphics[width=8cm]{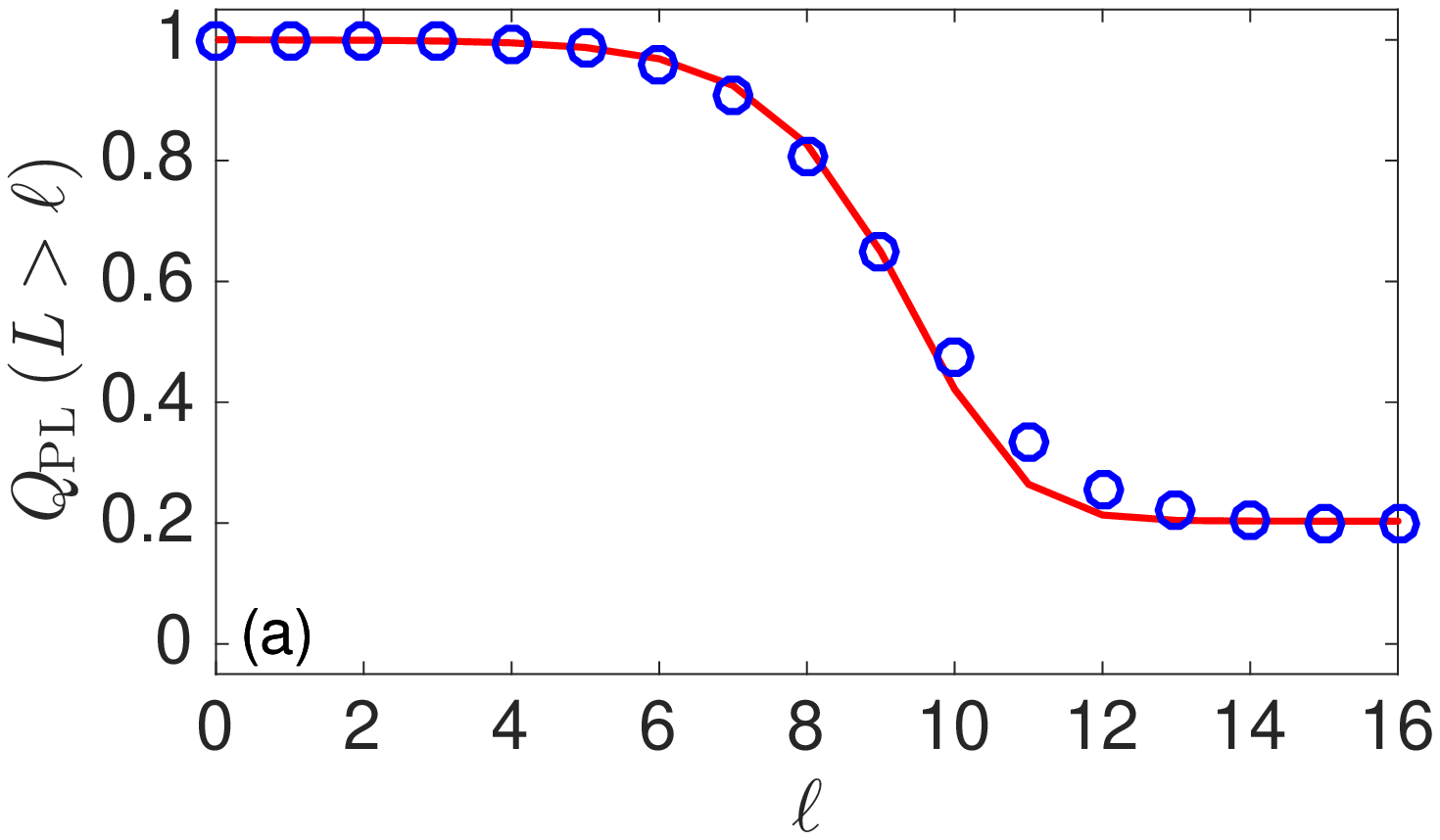} \\
\includegraphics[width=8cm]{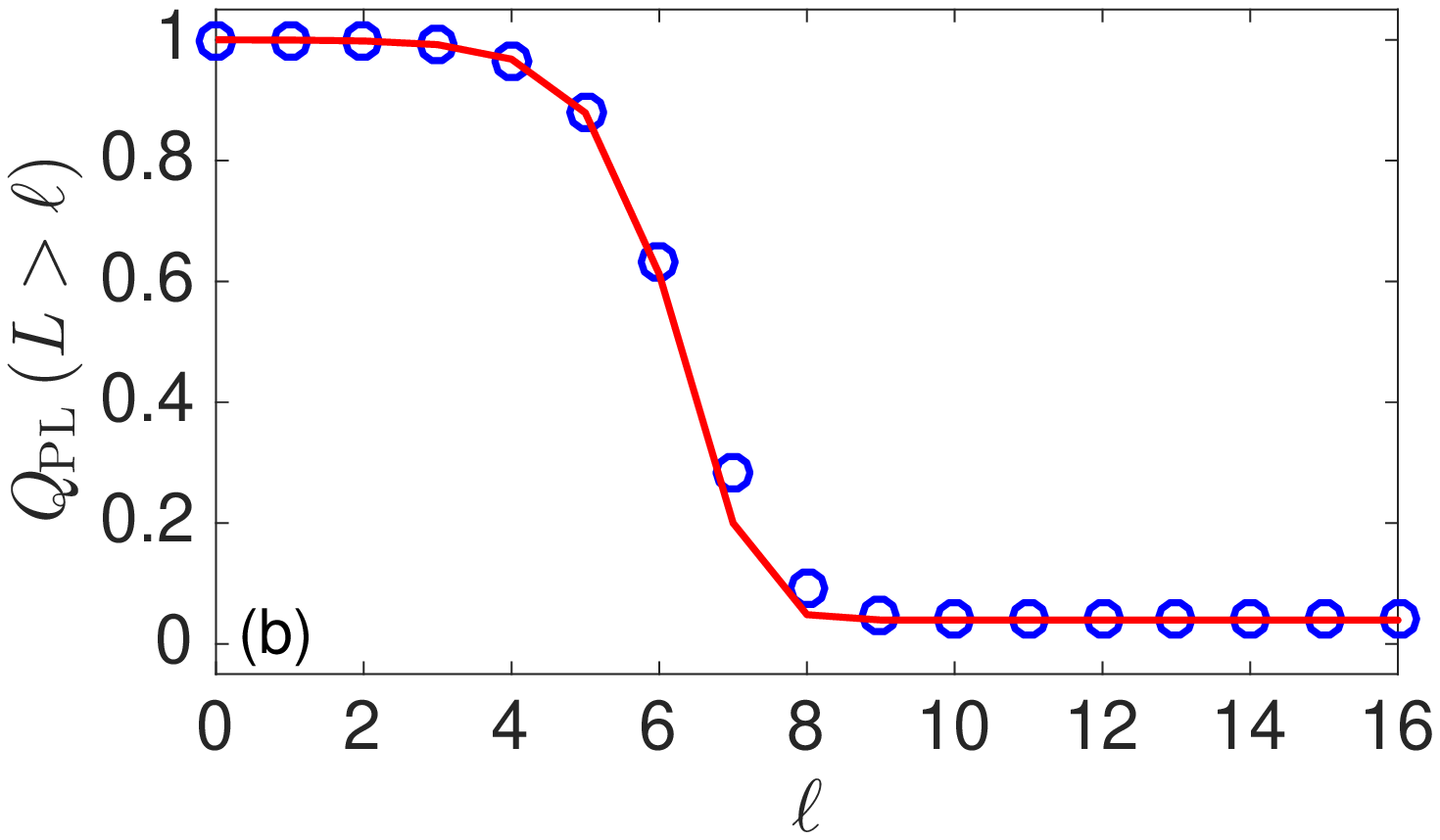} \\
\includegraphics[width=8cm]{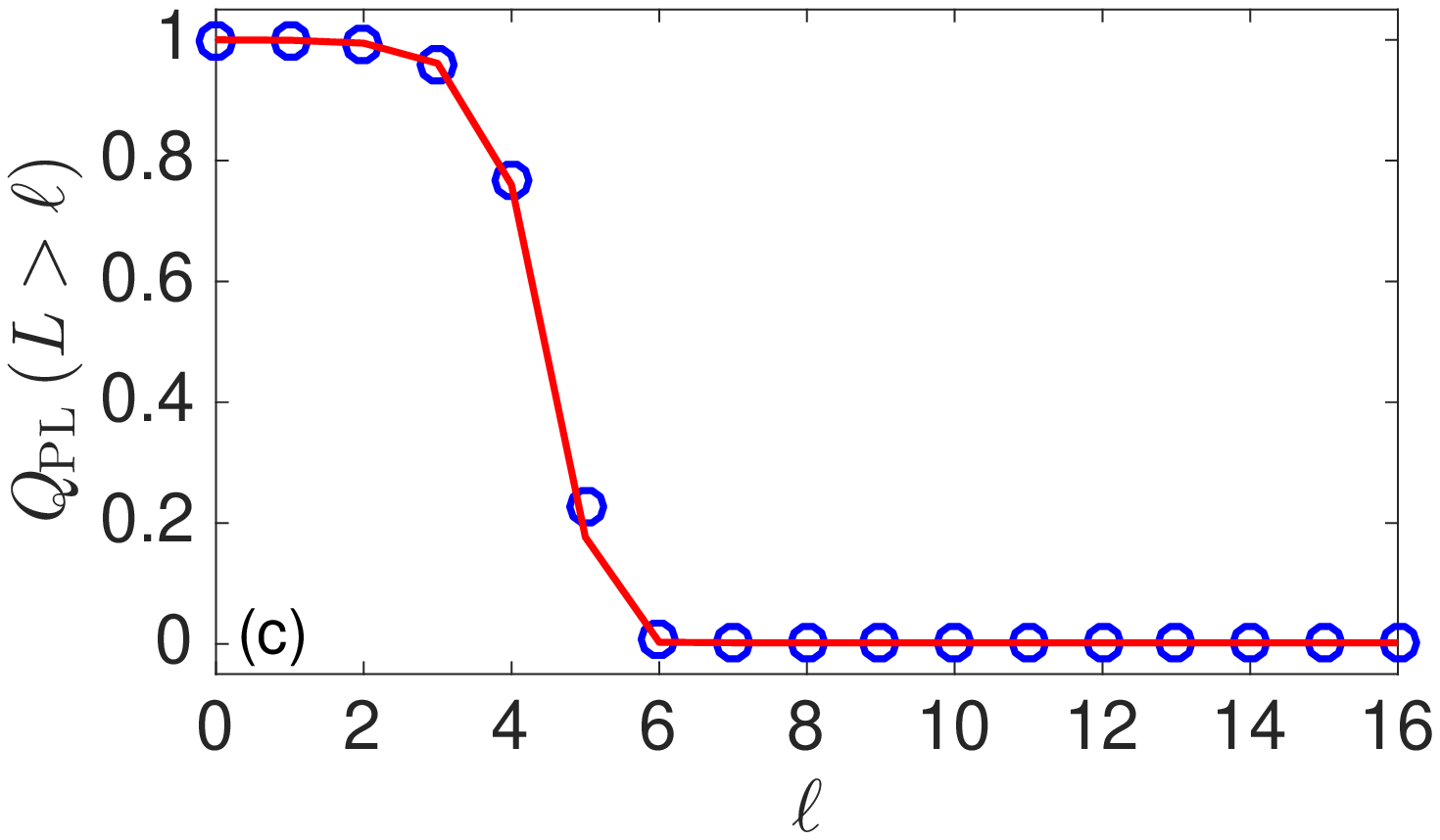}
\caption{
(Color online)
The tail distribution of shortest path lengths, 
$Q_{\rm PL}(L>\ell)$,  
for ER networks of $N=10^4$ nodes, 
and mean degree $c=2.5$ (a), $c=4$ (b) and $c=7$ (c).
The analytical results (solid lines),
obtained from Eq. (\ref{eq:Q}), 
are found to be in very good agreement
with the results of computer simulations (circles), 
which were averaged over 10 instances
of the network.
The tail distributions exhibit the characteristic shape of a monotonically
decreasing sigmoid function, with a non-zero asymptotic value
at large distances. 
The asymptotic value of 
$Q_{\rm PL}(L>\ell)$ 
at large distances accounts for the probability that
two randomly selected nodes do not reside on the same cluster.
As $c$ is increased, the inflection point,
which corresponds to the peak of the DSPL,
shifts to the left, namely distances in the network become shorter,
in agreement with the prediction of small-world theory.
Concurrently, the asymptotic tail moves down, 
reflecting the increasing size of the giant cluster.
}
\label{fig:3}
\end{figure}

In Fig. \ref{fig:3} 
we present the tail distributions
$Q_{\rm PL}(L>\ell)$,
for ER networks of $N=10^4$ nodes, 
with mean degree $c=2.5$ (a), $c=4$ (b) and $c=7$ (c).
The analytical results (solid lines),
obtained from Eq. (\ref{eq:Q}), 
are found to be in very good agreement
with the results of computer simulations (circles).
The tail distributions exhibit the characteristic shape of a monotonically
decreasing sigmoid function between two plateaus.
Their inflection points coincide with the peaks of the corresponding
probability distribution functions.
The tail distributions
$Q_{\rm PL}(L>\ell)$ 
exhibit non-zero asymptotic values
at large distances, which account for the probability that
two randomly selected nodes do not reside on the same cluster,
and thus the distance between them is $\ell = \infty$.
As $c$ is increased, the inflection point
shifts to the left, which means that distances in the network become shorter.
This can be understood in the framework of small-world theory, where
the mean distance is given by 
$\langle L \rangle \simeq \ln N / \ln c$.
Concurrently, the asymptotic value of
$Q_{\rm PL}(L>\ell)$ decreases,
due to the increasing size of the giant cluster.

\begin{figure}
\includegraphics[width=11cm]{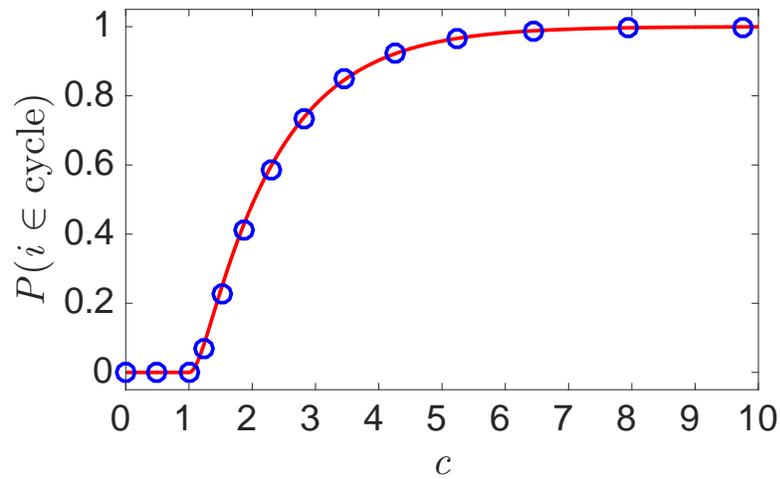}
\caption{
(Color online)
The probability $P(i \in {\rm cycle})$ 
that a random node, $i$, resides on
at least one cycle, versus the mean degree, 
$c$, in an ER network of $N=10^4$ nodes.
The analytical results (solid line),
obtained from Eq. (\ref{eq:ERincyc}),
are found to be in excellent agreement with the
results of computer simulations (circles).
}
\label{fig:4}
\end{figure}

Using Eqs. 
(\ref{eq:cyck})
and
(\ref{eq:cyc}),
the probability that a random node in an ER network
resides on at least one cycle can be expressed in the form

\begin{equation}
P(i \in {\rm cycle}) = 1 - \sum_{k=0}^{\infty}
\frac{e^{-c} c^k}{k!} (1-g^2)^{k(k-1)/2},
\label{eq:ERincyc}
\end{equation}

\noindent
where $g$ is given by Eq. (\ref{eq:g(c)}).
In Fig. \ref{fig:4} 
we present the probability
$P(i \in {\rm cycle})$
as a function of the mean degree, $c$,
for ER networks of $N=10^4$ nodes.
The analytical results (solid lines), 
obtained from Eq. 
(\ref{eq:ERincyc}),
are found to be in very good agreement 
with the results of computer simulation
(circles).
It is found that for $c<1$ there are no cycles
and thus 
$P(i \in {\rm cycle})=0$.
As $c$ is increased above $1$, the probability
$P(i \in {\rm cycle})$
increases sharply.

To obtain more accurate results, 
we consider a random node $i$ of a given degree, $k$,
and express the probability that it resides on at least
one cycle in the form

\begin{equation}
P(i \in {\rm cycle} |k) =
1 -
\sum_{k_1,k_2,\dots,k_k} \prod_{r=1}^k
%\frac{k_r}{\langle K \rangle} P(k_r)
\widetilde P(k_r)
\prod_{m<n} ( 1 - \tilde g_{k_m} \tilde g_{k_n} ).
\end{equation}

\noindent
In the case of an ER network,
where $P(k)$ is a Poisson distribution,
$\tilde g_k = g_k$
and
$k_r P(k_r)/\langle K \rangle = P(k_r-1)$,
where $k_r-1$ is the degree of the $r^{\rm th}$ 
neighbor
of node $i$ on the reduced network from which $i$ 
was removed.
Therefore, in the case of an ER network

\begin{equation}
P(i \in {\rm cycle} |k) =
1 -
\sum_{k_1,k_2,\dots,k_k} \prod_{r=1}^k
P(k_r)
\prod_{m<n} ( 1 - g_{k_m} g_{k_n} ).
\label{eq:cycle_k}
\end{equation}

\noindent
The evaluation of this product requires moments of $g_k$,
which can be expressed in a closed form as

\begin{equation}
\langle g_k^n \rangle =
\sum_{r=0}^{n} \binom{n}{r}
(-1)^r e^{-c[1 - (1-g)^r]}.
\label{eq:g_k^n}
\end{equation}

\noindent
The two lowest order moments are

\begin{equation}
\langle g_k \rangle = 1 - e^{-cg} = g,
\label{eq:<g_k>}
\end{equation}

\noindent
and

\begin{equation}
\langle g_k^2 \rangle = 1 - 2 e^{-cg} + e^{-cg(2-g)} 
= g^2 +(1-g)^2(e^{cg^2}-1).
\label{eq:<g_k^2>}
\end{equation}

\noindent
Inserting these moments into Eq. (\ref{eq:cycle_k})
we find that
the probability that a node of degree $k=2$ resides
on at least one cycle is 

\begin{equation}
P(i \in {\rm cycle} | K=2) = g^2.
\label{eq:icyc2}
\end{equation}

\noindent
Incidentally, this result coincides with the simpler form which
comes from Eq. (\ref{eq:cyck}).
For nodes of degree $k=3$

\begin{equation}
P(i \in {\rm cycle} | K=3) =
3g^2 - 3 g^2 \langle g_k^2 \rangle 
+ \langle g_k^2 \rangle^3.
\label{eq:icyc3}
\end{equation}

\noindent
At this order the result already deviates from those obtained
from the simpler approach of Eq. (\ref{eq:cyck}).
Analytical expressions for 
$P(i \in {\rm cycle} |K= k)$
with $k=4$ and $5$ are presented in Appendix C.

In Fig. \ref{fig:5} 
we present the conditional probability
$P(i \in {\rm cycle} |K= k)$, 
that a random node of degree $k$ resides on at least one cycle
as a function of the mean degree $c$,
for $k=2$ (a), $k=3$ (b) and $k=5$ (c).
The analytical results 
obtained from the
simpler approach of
Eq. (\ref{eq:cyck}) are shown in dashed lines.
The analytical results obtained from 
the more detailed approach of
Eq. (\ref{eq:Pncc2}) 
are given explicitly in Eqs. 
(\ref{eq:icyc2}), (\ref{eq:icyc3}),
(\ref{eq:icyc4}) and (\ref{eq:icyc5}).
These results are shown in solid lines.
Incidentally, the two analytical curves coincide for $k=2$,
while for $k=3$ and $5$, the more detailed theory is found to
be in a better agreement with the
results of computer simulations (circles).

The DSCL of an ER network is given by

\begin{equation}
P_{\rm CL}(L>\ell) = (1+c) e^{-c} 
+ \sum_{k=2}^{\infty} \frac{e^{-c} c^k}{k!} {Q}_{\rm PL}(L>\ell-2)^{\binom{k}{2}}.
\label{eq:ERdscl}
\end{equation}

\begin{figure}
\includegraphics[width=9cm]{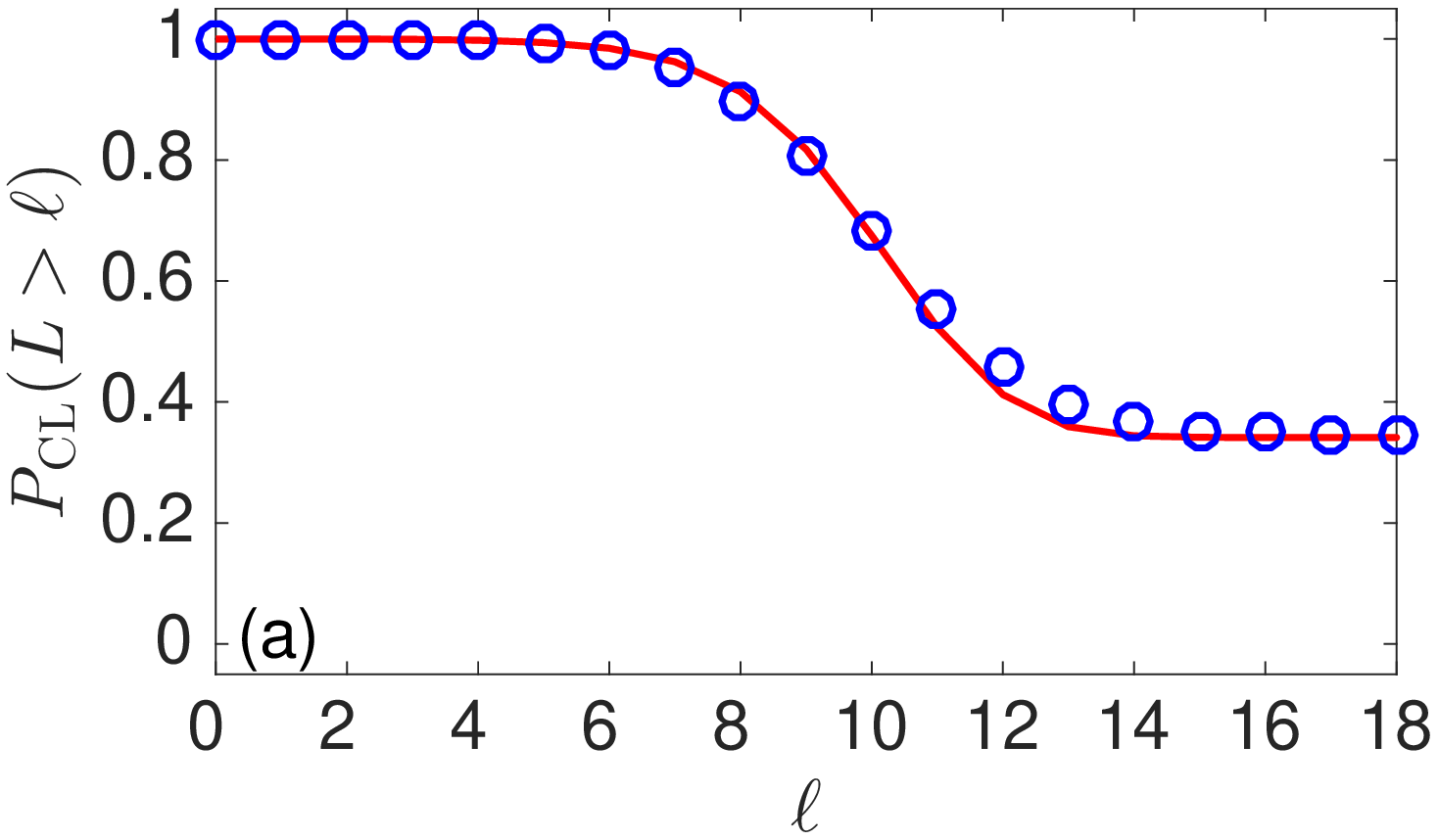}
\includegraphics[width=9cm]{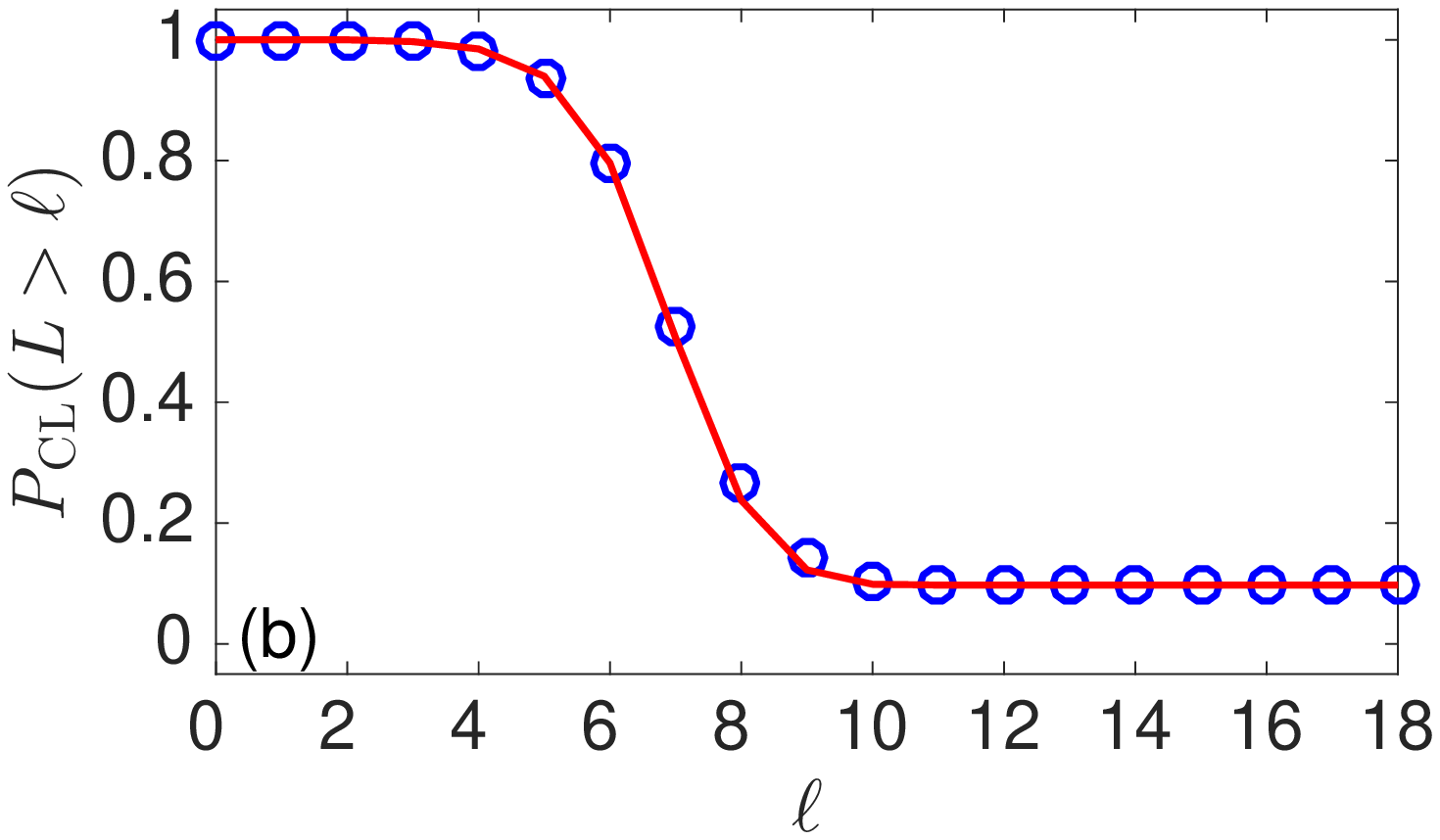}
\includegraphics[width=9cm]{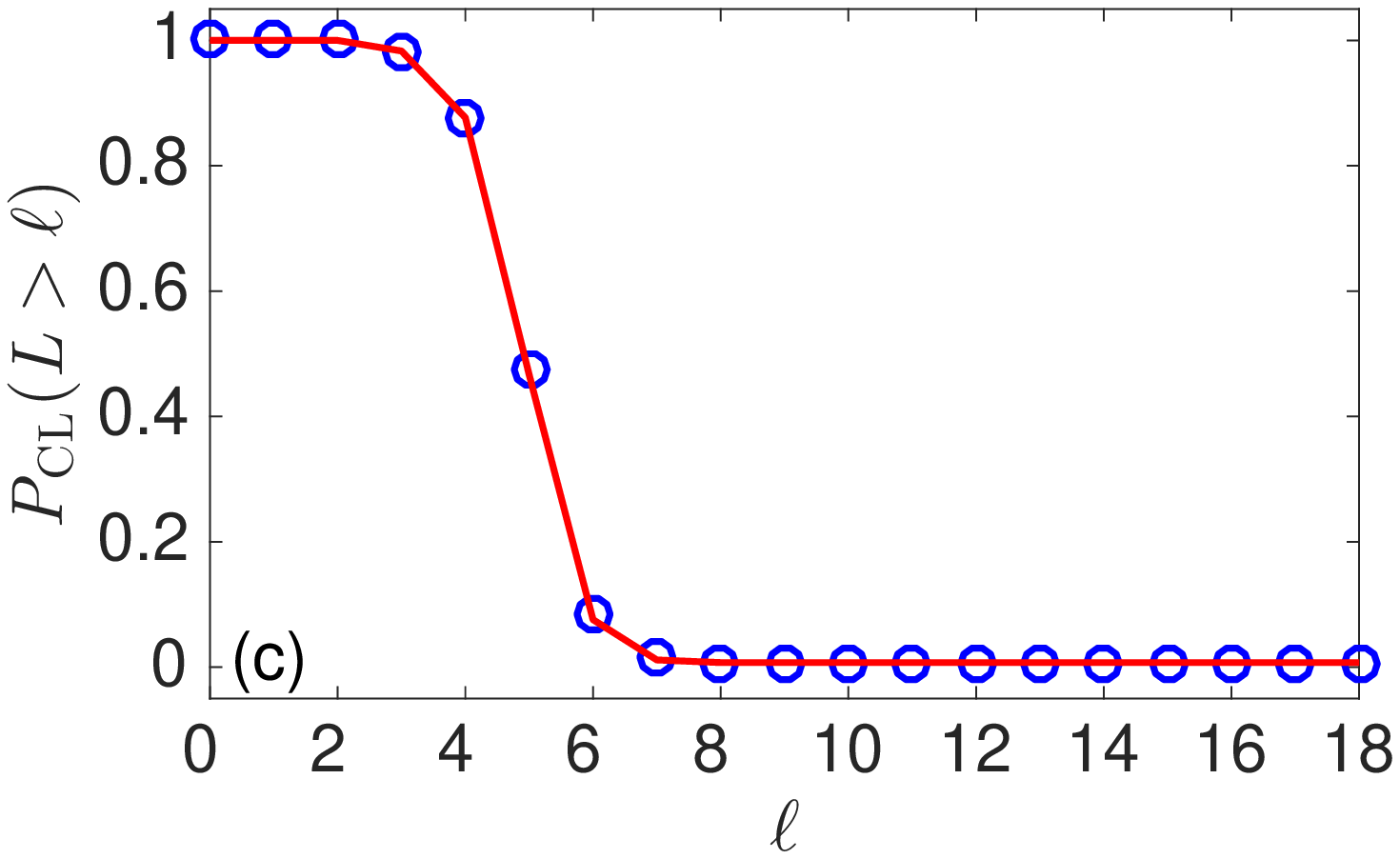}
\caption{
(Color online)
The tail distribution of shortest cycle lengths, $P_{\rm CL}(L>\ell)$,
for ER networks of $N=10^{4}$ nodes,
where $c=2.5$ (a), $c=4$ (b) and $c=7$ (c).
The analytical results (solid lines),
obtained from Eq. (\ref{eq:ERdscl}),
are found to be in very good agreement with the results of computer 
simulations (circles).
}
\label{fig:6}
\end{figure}

In Fig. \ref{fig:6} we present the
tail distributions 
$P_{\rm CL}(L>\ell)$
for ER networks of $N=10^{4}$ nodes,
where $c=2.5$ (a), $c=4$ (b) and $c=7$ (c).
The analytical results (solid lines), 
obtained from Eq. (\ref{eq:ERdscl}),
are in good agreement with computer 
simulations (circles).
The tail distribution exhibits a monotonically decreasing sigmoid shape
from the $P_{\rm CL}(L>\ell)=1$ plateau on the left to 
$P_{\rm CL}(L>\ell)=P(i \notin {\rm cycle})$ on the right,
since the
height of the second plateau represents the fraction of nodes
which do not reside on any cycle.
This fraction decreases as the mean degree, $c$, is increased,
namely the probability that a random node resides on at least one
cycle increases as $c$ is increased.
The peak of the corresponding probability distribution function,
$P_{\rm CL}(L=\ell)$,
shifts to the left as $c$ is increased.
These results imply that as the network becomes more strongly 
connected the shortest cycles become more numerous and shorter.

\begin{figure}
\includegraphics[width=9cm]{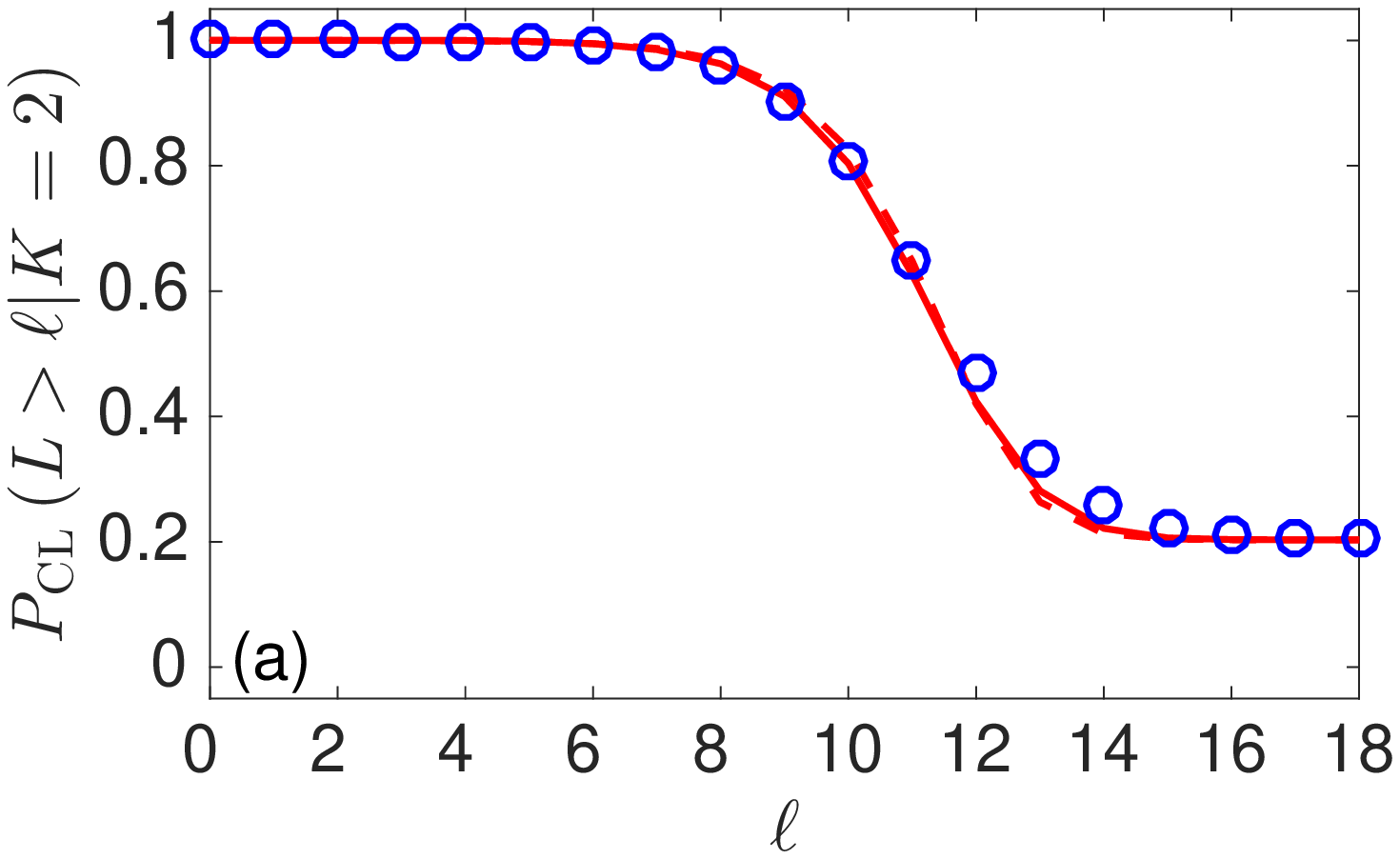}
\includegraphics[width=9cm]{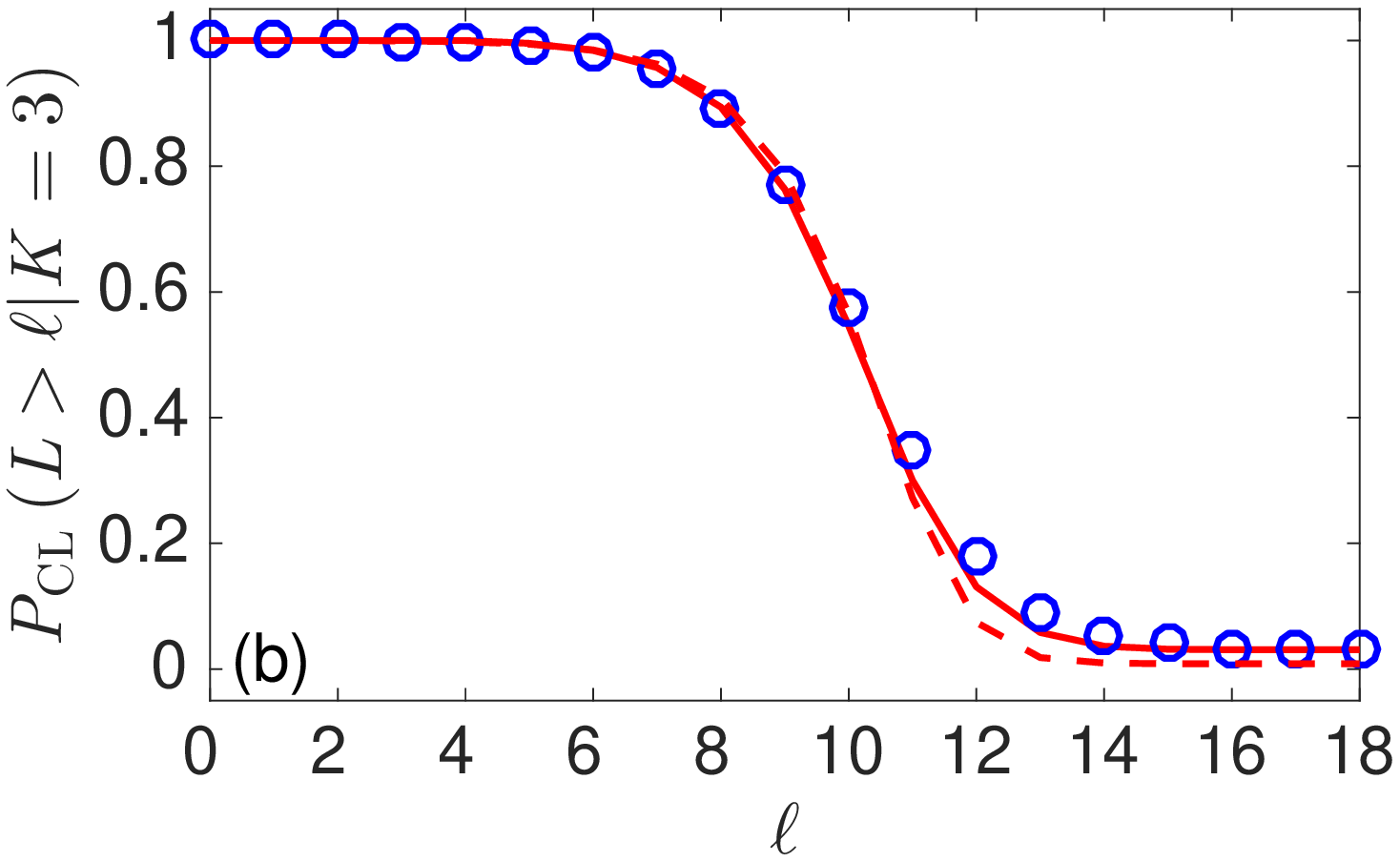}
\includegraphics[width=9cm]{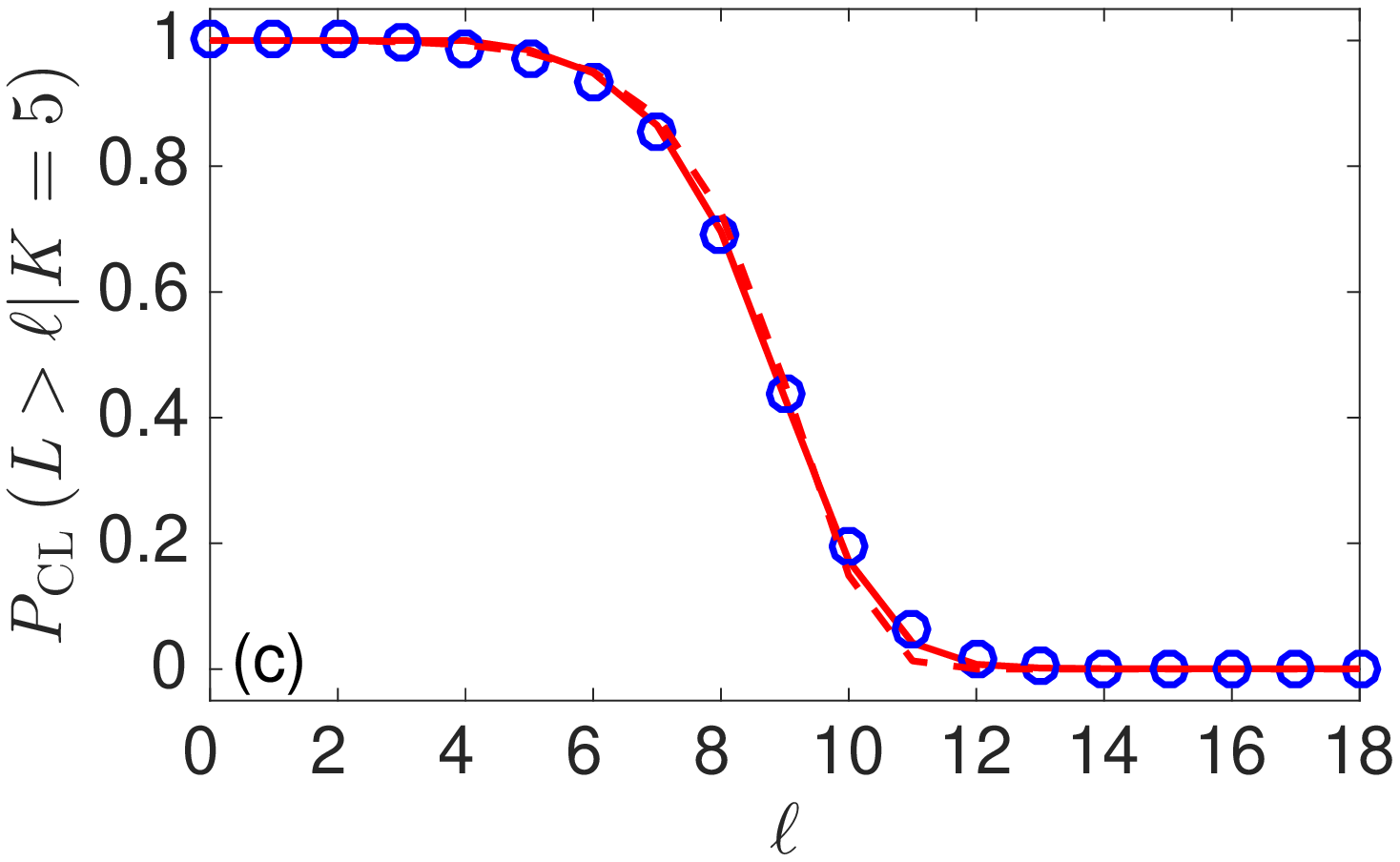}
\caption{
(Color online)
The conditional distribution of shortest cycle lengths,
$P_{\rm CL}(L>\ell | K=k)$,
for ER networks of $N=10^{4}$ nodes,
where $c=2.5$ and $k=2$ (a), $k=3$ (b) and $k=5$ (c).
The analytical results 
obtained from the
simpler approach of
Eq. (\ref{eq:dscl_c1}) are shown in dashed lines,
while the analytical results obtained from 
the more detailed approach of
Eq. (\ref{eq:dscl_c2}) are
shown in solid lines.
The more detailed theory is found to
be in better agreement with the
results of computer simulations (circles).
}
\label{fig:7}
\end{figure}

In Fig. \ref{fig:7} we present the conditional tail distribution
$P_{\rm CL}(L>\ell |K= k)$
for an ER network of $N=10^4$ nodes and $c=2.5$,
where $k=2$ (a), $k=3$ (b) and $k=5$ (c).
The analytical results 
obtained from the
simpler approach of
Eq. (\ref{eq:dscl_c1}) 
are shown in dashed lines,
while the analytical results obtained from 
the more detailed approach of
Eq. (\ref{eq:dscl_c2}) are
shown in solid lines.
The two analytical curves are almost indistinguishable for $k=2$,
and are both in very good agreement with the results of 
computer simulations (circles).
For $k=3$ and $5$, the more detailed theory provides a
better agreement with the
results of computer simulations (circles).

The conditional tail distribution retains the qualitative features of the
sigmoid shape. 
The asymptotic value at large $\ell$ is
$P_{\rm CL}(L>\ell)=P(i \notin {\rm cycle})$, which decreases as $k$ is 
increased, which means that the probability that a random node
of degree $k$ resides on at least one cycle increases as $k$ is
increased. 
The peak of the corresponding probability distribution function,
$P_{\rm CL}(L=\ell | K=k)$,
shifts to the left as $k$ is increased,
which means that for node of higher degree the shortest cycle
is shorter.

The probability that a random node, $i$, of degree $k$ resides on at least one cycle
is a monotonically increasing function of $k$.
The length $\ell$ of the shortest cycle tends to decrease as a function
of $k$. This is due to the fact that the length of the shortest cycle
is determined by the shortest path among all the paths connecting
neighbors of $i$, and the number of such pairs increases quadratically with $k$. 

\begin{figure}
\includegraphics[width=10cm]{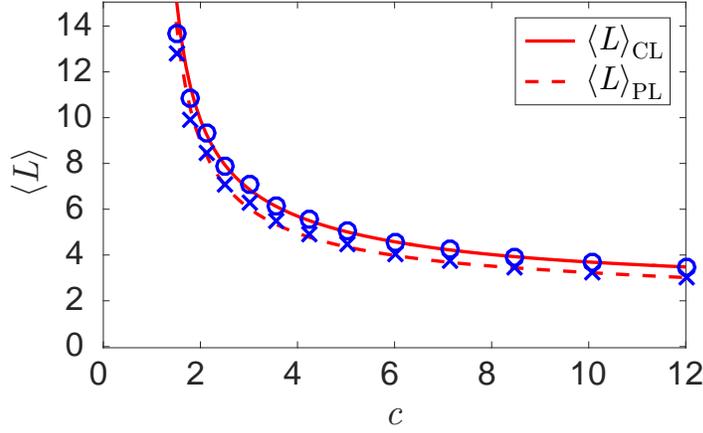}
\caption{
(Color online)
Analytical results for the
mean, 
$\langle L \rangle_{\rm CL}$, 
of the distribution of shortest cycle lengths (solid line), 
and for the mean,
$\langle L \rangle_{\rm PL}$,
of the distribution of shortest path lengths (dashed line), 
as a function of the mean degree, $c$, for ER networks 
of $N=10^3$ nodes.
Both curves are found to be in very good agreement with
the corresponding results obtained from computer simulations 
(circles for 
$\langle L \rangle_{\rm CL}$
and $\times$ for
$\langle L \rangle_{\rm PL}$).
It is found that
$\langle L \rangle_{\rm CL}$
is slightly larger than
$\langle L \rangle_{\rm PL}$
for all values of $c$.
}
\label{fig:8}
\end{figure}

In Fig. \ref{fig:8} we present
analytical results for the
mean, $\langle L \rangle_{\rm CL}$, of the DSCL 
as a function of the mean degree, $c$, for ER networks 
of $N=10^3$ nodes (solid line).
The results are in very good agreement with
computer simulations (circles).
The mean, $\langle L \rangle_{\rm CL}$ is a monotonically
decreasing function of $c$. 
It exhibits a sharp decrease in the dilute network limit,
which becomes more moderate as the network becomes
more dense.
For comparison, we also present 
analytical (dashed line) and numerical ($\times$) results for
the mean, 
$\langle L \rangle_{\rm PL}$,
of the DSPL as a function of $c$
(dashed line).
It is found that for the entire range of values of $c$, the
mean of the DSCL is slightly larger than the mean of the
corresponding DSPL.
This can be understood as follows. 
The length of the shortest cycle on which a random node, $i$, resides, 
consists of the shortest path between a pair of its neighbors, plus $2$ 
for the two edges connecting $i$ to these neighbors. 
This suggests that $\langle L \rangle_{\rm CL}$ should be longer
by about two units than $\langle L \rangle_{\rm PL}$.
However, the shortest path between neighbors of $i$ which is
incorporated in the shortest cycle is the shortest among the 
shortest paths connecting {\it all} pairs of neighbors of $i$. Thus,
it tends to be shorter than the path between two random nodes.
As a result, the difference
$\Delta = \langle L \rangle_{\rm CL} - \langle L \rangle_{\rm PL}$
is smaller than $2$.

\begin{figure}
\includegraphics[width=10cm]{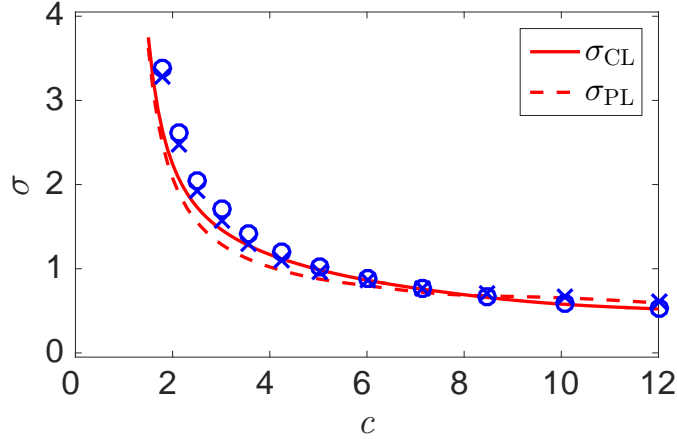}
\caption{
(Color online)
Analytical results for the
standard deviations, 
$\sigma_{\rm CL}$, 
of the distribution of shortest cycle lengths (solid line), 
and
$\sigma_{\rm PL}$ 
of the distribution of shortest path lengths (dashed line),
as a function of the mean degree, $c$, for ER networks 
of $N=10^3$ nodes.
For small values of $c$, the analytical results appear to under-estimate
the standard deviations with respect to the results of computer simulations for
$\sigma_{\rm CL}$ (circles)
and
$\sigma_{\rm PL}$ ($\times$).
}
\label{fig:9}
\end{figure}

In Fig. \ref{fig:9} we present
the standard deviation of the DSCL, $\sigma_{\rm CL}$ 
as a function of the mean degree, $c$, for ER networks 
of $N=10^3$ nodes.
For small values of $c$,
the analytical results (solid line) 
under-estimate the standard deviation, as can be seen
from the comparison with
the results of computer simulations (circles).
We also show the
analytical (dashed line) and numerical ($\times$) results for the
standard deviation of the DSPL,
$\sigma_{\rm PL}$, for the same networks,
which exhibits the same qualitative features.

\subsection{Random regular graphs}

In a random regular graph with $c \ge 3$
the giant cluster encompasses the whole network.
Therefore, $g = {\widetilde g}=1$
(for more details see Appendix D).
Moreover, in this case the DSPLs and the overall DSPLs
are identical since all pairs of nodes reside on the
giant cluster. 
The generating functions for the random regular graph are given
by 
Eqs. (\ref{eq:g1rrg}) and (\ref{eq:g0rrg}).
The DSCL is given by

\begin{equation}
P_{\rm CL} (L>\ell)
= 
\left[ \widehat P_{\rm PL} (L>\ell-2) \right]^{\binom{c}{2}}.
\label{eq:DSCLr1}
\end{equation}

\noindent
In order to proceed we shall first calculate the
conditional probabilities 
$\widehat P_{\rm PL}(L>\ell | L>\ell-1)$ 
using 
the recursion equation (\ref{eq:P_rec2sh}) and the initial 
condition (\ref{eq:L1hat}). 
This yields

\begin{equation}
\widehat P_{\rm PL}(L>\ell|L>\ell-1) = \left[ 1-\frac{(c-1)^2}{(N-1)c} \right]^{(c-1)^{\ell-1}}.
\end{equation}

\noindent
Assuming that the size of the network is large $N \gg 1$,
we can approximate the above to

\begin{equation}
\ln \widehat P_{\rm PL}(L>\ell|L>\ell-1) \simeq -\frac{(c-1)^{\ell+1}}{c N} 
+ {\mathcal O} \left(\frac{1}{N^2}  \right).
\end{equation}

\noindent
By inserting the conditional distribution into 
Eq. (\ref{eq:prodh}) we can obtain the tail distribution 

\begin{equation}
\widehat P_{\rm PL}(L>\ell) \simeq 
\exp \left[ -\frac{(c-1)^2}{cN} \frac{(c-1)^{\ell}-1}{c-2} \right].
\end{equation}

\noindent
We can use this DSPL inside Eq. (\ref{eq:DSCLr1}), to get

\begin{equation}
P_{\rm CL}(L>\ell) \simeq 
\exp \left[ -\frac{(c-1)^3}{2N} \frac{(c-1)^{\ell-2}-1}{c-2} \right].
\label{eq:DSCLr2}
\end{equation}

\begin{figure}
\includegraphics[width=9cm]{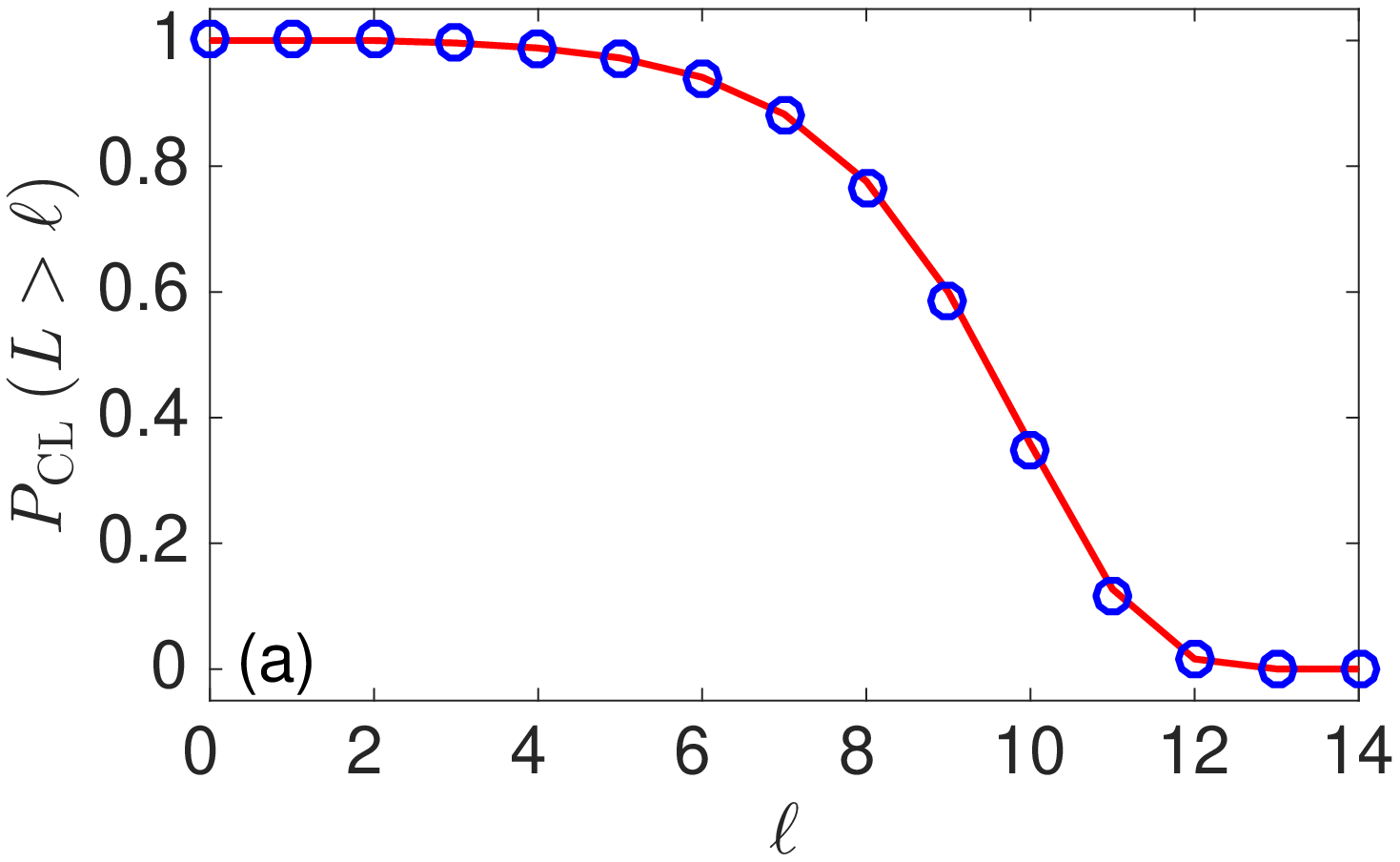}
\includegraphics[width=9cm]{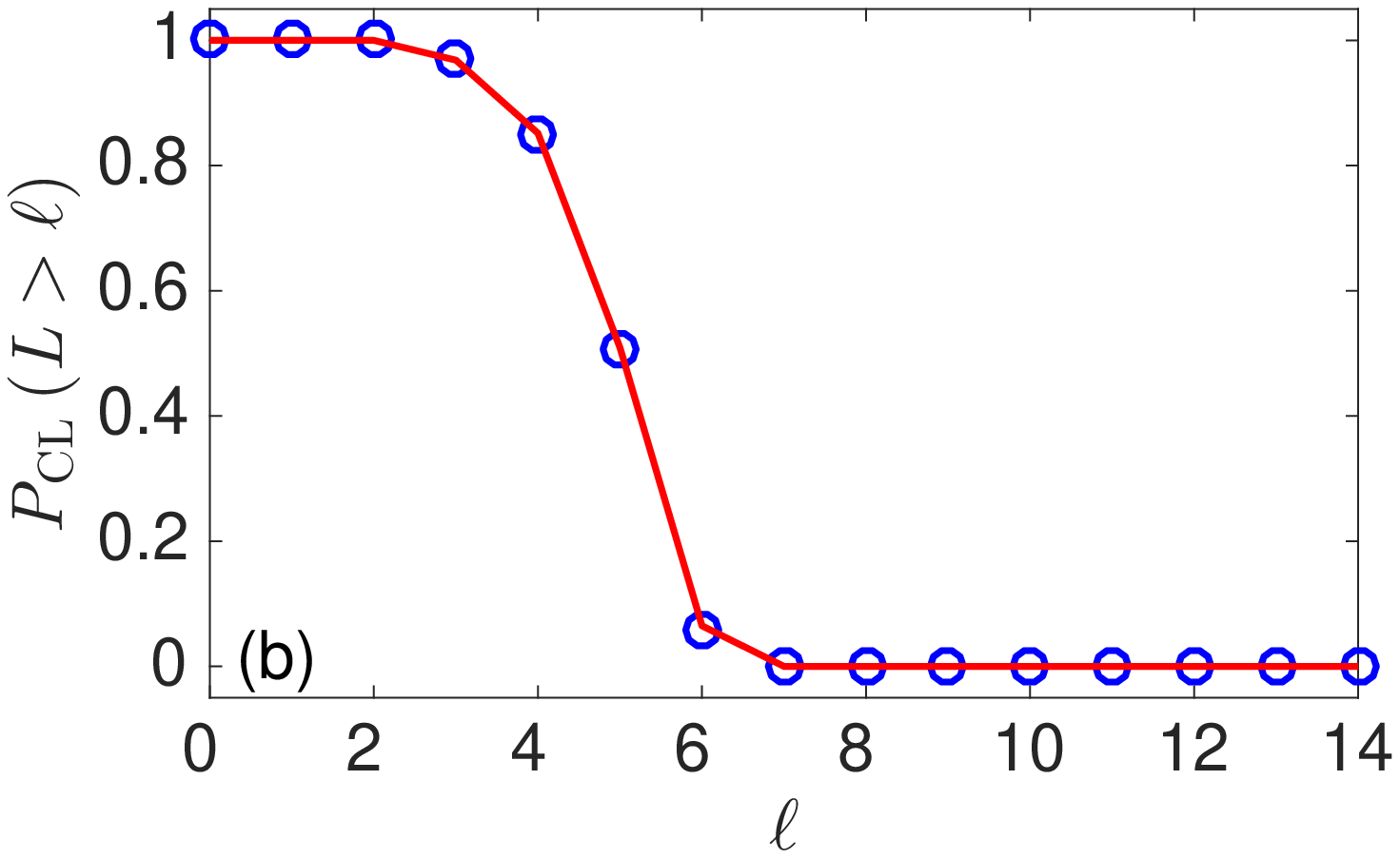}
\includegraphics[width=9cm]{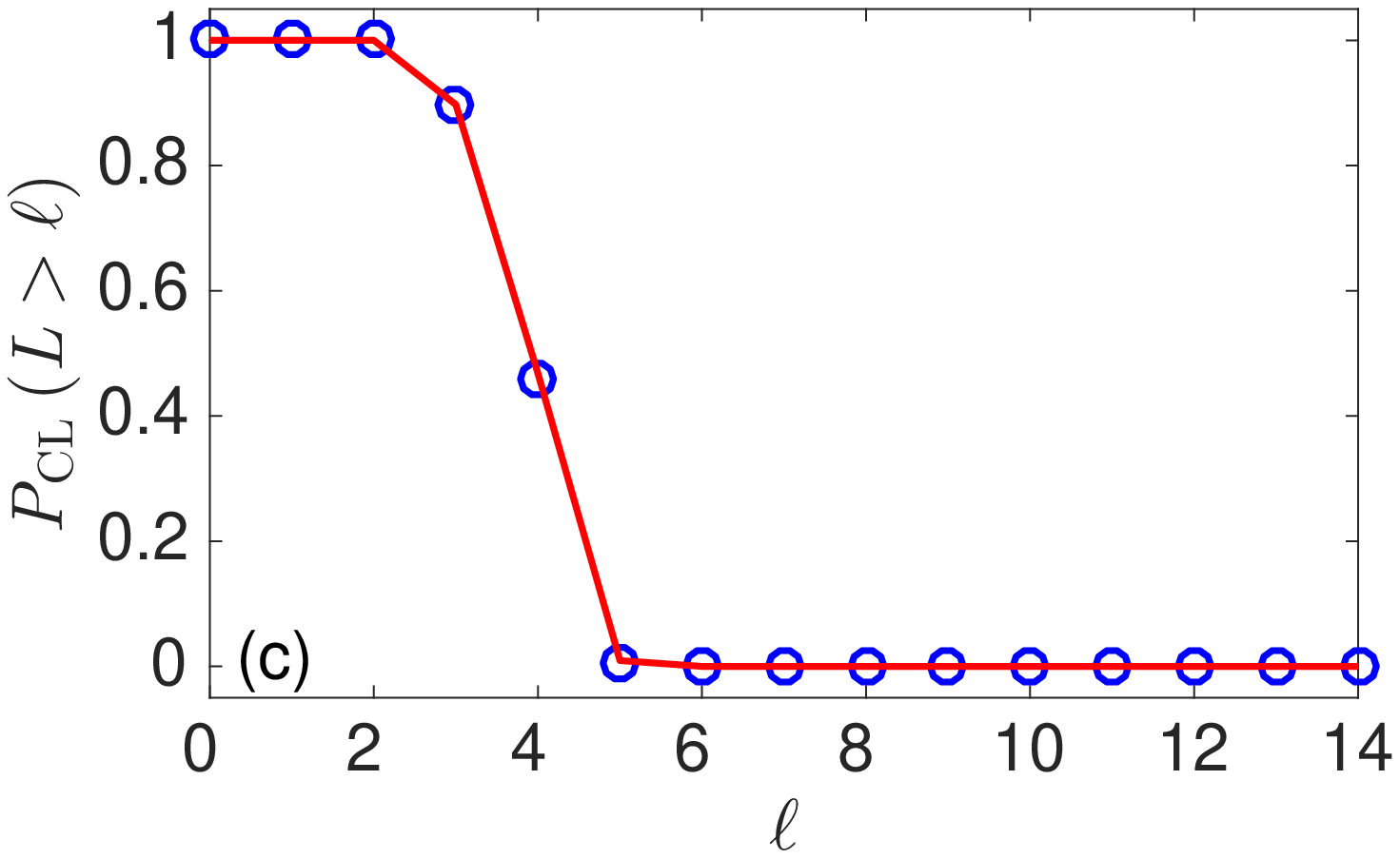}
\caption{
(Color online)
The tail distributions of shortest cycle lengths, 
$P_{\rm CL}(L>\ell)$,
for random regular graphs of $N=10^{3}$ nodes and 
$c=3$ (a), $c=5$ (b) and $c=7$ (c).
The analytical results (solid line),
obtained from Eq. (\ref{eq:DSCLr2}),
are found to be in excellent agreement with the results of 
computer simulations (circles).
}
\label{fig:10}
\end{figure}

\noindent
In Fig. \ref{fig:10} we present the
DSCL for random regular graphs of $N=10^{3}$ nodes with $c=3$ (a),
$c=5$ (b) and $c=7$ (c).
The analytical results (solid lines),
obtained from Eq. (\ref{eq:DSCLr2}),
are found to be 
in excellent agreement with the results of 
computer simulations (circles).
Since Eq. (\ref{eq:DSCLr2}) is based on exact results for the 
DSPL, we conjecture that it is an exact result for the DSCL of the
random regular graph.

\subsection{Scale free networks}

Consider a configuration model network with a power-law degree distribution, $P(k)$,
given by 

\begin{equation}
P(k) = \frac{ k^{-\gamma} }{ \zeta(\gamma,k_{\rm min}) 
- \zeta(\gamma,k_{\rm max}+1) },
\label{eq:PLnorm}
\end{equation}

\noindent
where 
$k_{\rm min} \le k \le k_{\rm max}$
and
$\zeta(s,a)$ is the Hurwitz zeta function 
\cite{Olver2010}.
Here we focus on the case in which $\gamma > 2$, in which the
mean degree $\langle K \rangle$ is bounded even for 
$k_{\rm max} \rightarrow \infty$.
We further restrict our analysis to the case in which $k_{\rm min} \ge 2$,
namely the network does not include isolated nodes and leaf nodes.
Under these conditions 
$g = \tilde g = 1$, namely the giant cluster encompasses the entire network
(for more details see Appendix E).
As a result, $\widehat Q_{\rm PL}(L>\ell) = \widehat P_{\rm PL}(L>\ell)$.
Thus, the DSCL can be expressed in the form

\begin{equation}
P_{\rm CL}(L > \ell) = \sum_{k=2}^{\infty} 
P(k) \widehat P_{\rm PL}(L>\ell-2)^{\binom{k}{2}},
\end{equation}

\noindent
where $\widehat P_{\rm PL}(L>\ell-2)$ is calculated using 
Eqs. (\ref{eq:L1hat})-(\ref{eq:prodh}),
where $\langle K \rangle$ and $\langle K^2 \rangle$
are given by Eqs. (\ref{eq:Kmsf}) and (\ref{eq:K2msf}),
respectively.

\begin{figure}
\includegraphics[width=9cm]{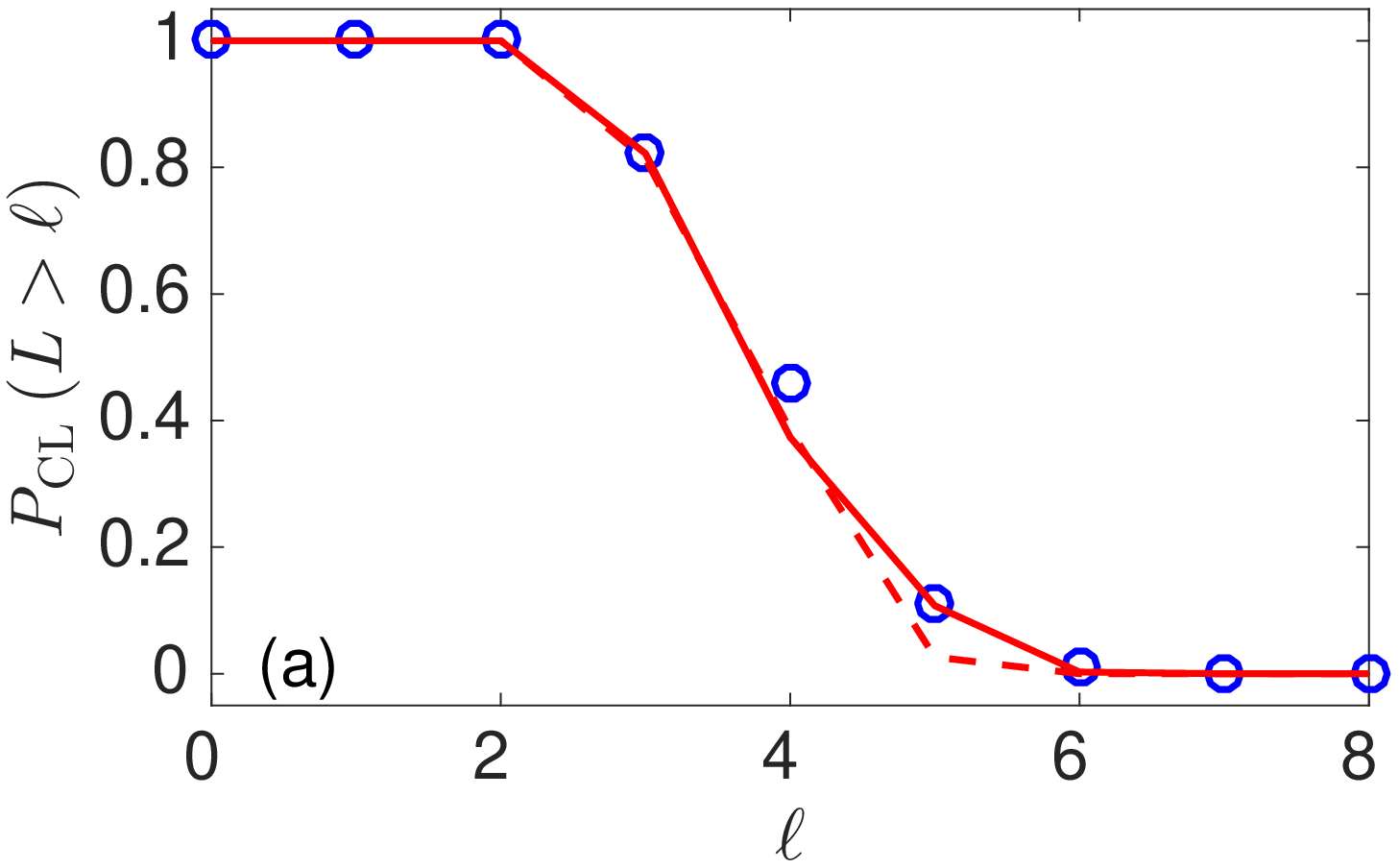}
\includegraphics[width=9cm]{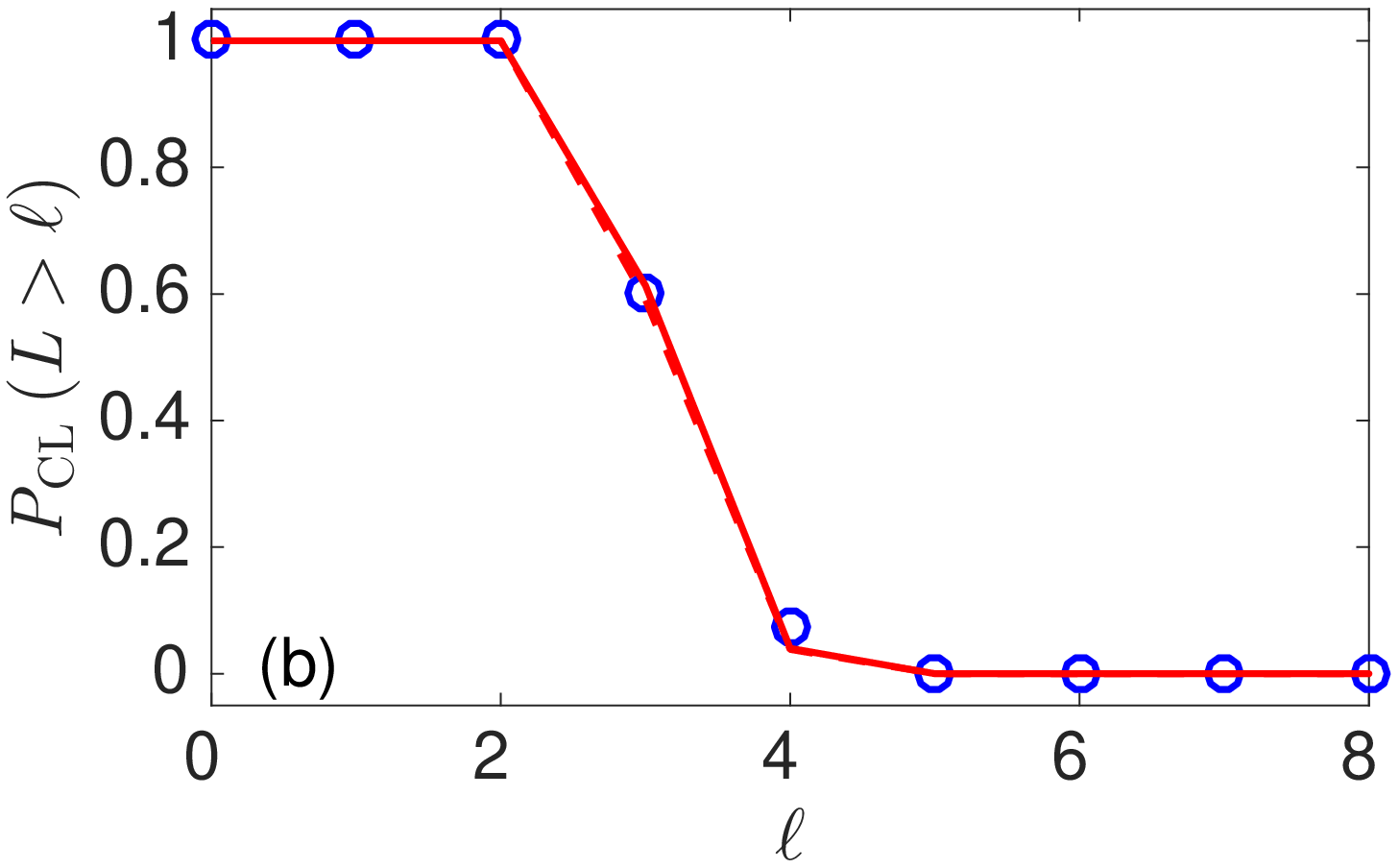}
\includegraphics[width=9cm]{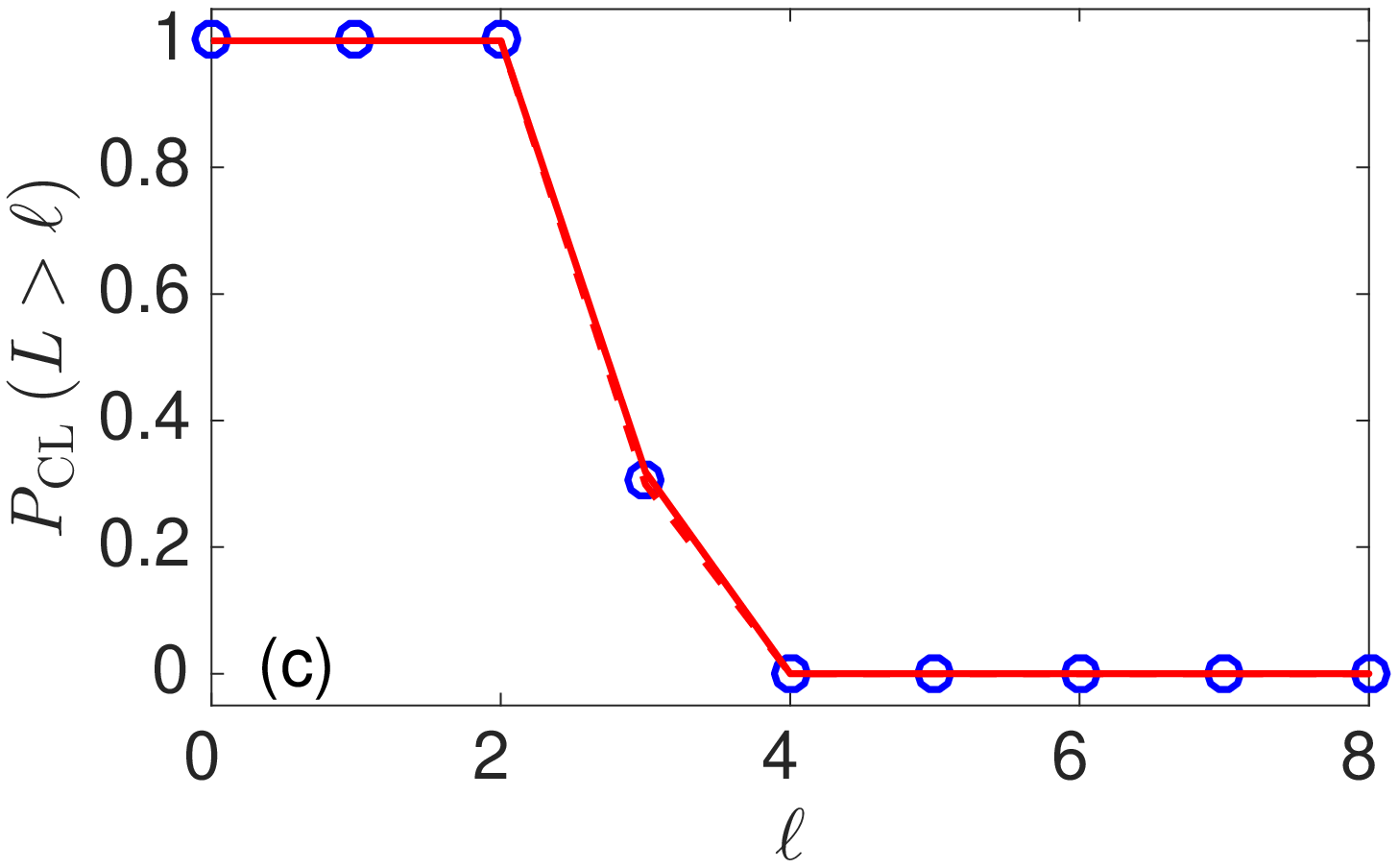}
\caption{
(Color online)
The tail distribution of shortest cycle lengths, 
$P_{\rm CL}(L>\ell)$,
for scale free networks of $N=10^{3}$ nodes and a 
power-law degree distribution with
$\gamma=2.5$ and $k_{\rm min}=3$ (a), $5$ (b) and $8$ (c).
The analytical results 
obtained from the
simpler approach of
Eq. (\ref{eq:dscl_c1}) 
are shown in dashed lines,
while the analytical results obtained from 
the more detailed approach of
Eq. (\ref{eq:dscl_c2}) 
are shown in solid lines.
Both results are in very good agreement
with the results of computer simulations (circles), 
except for some deviation of the simpler approach 
for $k{\rm min}=3$.
}
\label{fig:11}
\end{figure}

In Fig. \ref{fig:11} we present the
tail distribution $P_{\rm CL}(L>\ell)$,
for a configuration model network of $N=10^{3}$ nodes and a 
power-law degree distribution with
$\gamma=2.5$ and $k_{\rm min}=3$ (a), $5$ (b) and $8$ (c).
The analytical results
obtained from the
simpler approach of
Eq. (\ref{eq:dscl_c1}) 
are shown in dashed lines,
while the analytical results obtained from 
the more detailed approach of
Eq. (\ref{eq:dscl_c2}) 
are shown in solid lines.
The results of the more detailed approach were obtained from 
$10^4$ Monte Carlo samplings of the degrees $k,k_1,k_2,\dots,k_k$.
Both results are found to be in very good agreement with the results of 
computer simulations (circles), except for one data point of the simpler 
approach, for $k_{\rm min}=3$, at $\ell=5$, which is significantly 
lower than the simulation result.
It is observed that as $k_{\rm min}$ is increased, the distances
in the network become shorter.

\section{Discussion}

An important distinction in network theory is between networks
which exhibit a tree structure and networks which include cycles.
In network growth models, the existence of cycles is determined
by the growth rules of the network. For example, in the 
Barab\'asi-Albert model
\cite{Barabasi1999,Albert2002},
the existence of cycles depends
on the number of nodes, $m$, which are added at each time step. 
In the case in which $m=1$, the model gives rise to a stochastic tree structure
\cite{Drmota1997,Drmota2005}, 
while for $m \ge 2$ it forms cycles.

In equilibrium networks such as
configuration model networks, one can distinguish between three
situations, which are determined by the degree distribution $P(k)$.
In the sub-percolation regime of dilute networks, the network 
consists of finite tree components, whose size does not scale with $N$.
In this regime, the number of cycles does not scale with $N$.
Above percolation, the network consists of a giant cluster,
which includes cycles, in addition to many finite components.
As the network becomes denser, the number of cycles increases
and their typical length becomes shorter.
In the regime of dense networks, the giant cluster encompasses the
entire network and there are many short cycles.

The degree distribution plays a crucial role in shaping the properties
of cycles in a network. In particular, isolated nodes (of degree $k=0$)
and leaf nodes (of degree $k=1$) cannot reside on any cycle.
Only nodes of degrees $k \ge 2$ may reside on a cycle.
Still, some nodes of degrees $k \ge 2$ do not reside on 
any cycle. Instead, they reside on a tree component which can be
either isolated or connected to the giant cluster.

There are interesting connections between the DSCL and the DSPL
of a configuration model network.
For a random node, $i$, the cycles on which it resides consist
of paths between pairs of neighbors of $i$ and two edges from
$i$ to these neighbors.
The shortest cycle length is thus given by the shortest path
between all pairs of neighbors of $i$ plus $2$.
A naive expectation would thus be that the shortest
cycles are longer than the shortest paths by $2$ units.
From Fig. \ref{fig:8} we observe that the mean cycle length
is longer than the mean path length by about one unit over
a broad range of values of $c$ in the ER network.
To understand this point, we recall that the shortest cycle
on which a random node $i$ of degree $k$ resides, is composed of
the shortest path among all the $\binom{k}{2}$ paths connecting
pairs of neighbors of $i$. 
Another issue is the fact that the degrees of the 
neighbors of $i$ are not uniformly sampled from $P(k)$ but
from $\widetilde P(k)$.
The mean path length between pairs of neighbors of $i$ is
given by $\langle {\widehat L} \rangle$, while the mean
path length between pairs of random nodes is given by
$\langle L \rangle$.
Clearly, the path lengths between nodes of higher degrees
are shorter than between nodes of lower degrees, as
can be seen from Eqs. 
(\ref{eq:P})-(\ref{eq:whP}).
It is thus interesting to compare the mean degrees of
$P(k)$ and $\widetilde P(k)$.
The former is given by $\langle K \rangle$ while the
latter is $\langle K^2 \rangle / \langle K \rangle$.
In our context, the effective degree of a neighbor of a random node $i$
is given by the connective constant 
$\mu = \langle K^2 \rangle / \langle K \rangle - 1$,
where the edge connecting $i$ and its neighbor is removed.
It turns out that $\mu$ may be larger than, equal to or smaller than
$\langle K \rangle$ in different network ensembles. 
In the case of the ER ensemble, a special symmetry gives rise to
$\mu = \langle K \rangle$.
In the random regular graph, it turns out that $\mu = c-1$ and 
thus $\mu < \langle K \rangle$.
In configuration models with a power-law degree distribution
and $2 < \gamma \le 3$,
the moment
$\langle K^2 \rangle$ diverges and
thus 
$\mu > \langle K \rangle$.
In those cases in which 
$\mu > \langle K \rangle$,
the mean distance between neighbors of $i$ 
is smaller than the mean distance between random nodes,
and vice versa.
Therefore, the difference between the mean of the DSCL and 
the mean of the DSPL is determined by a combination of these
conflicting effects.

The results presented above have implications for the stability of configuration
model networks to node deletion processes due to failures or attacks.
In particular, if a node of degree $k \ge 2$, which does not reside on any cycle,
is removed, the network breaks down to $k$ separate components. 
Thus, nodes of degree $k \ge 2$ which do not reside on any cycle are
articulation points
\cite{Tian2017}.

In this paper we have studied configuration model networks in
which the DSCL is completely determined by the degree distribution
$P(k)$. Recently, other network ensembles were introduced, which
include many short cycles, where the cycle lengths are controlled by
various constraints
\cite{Roberts2014,Coolen2016}.
It would be interesting to generalize the calculation of the 
DSCL to such networks.

Knowing the properties of cycles is important for the study of
many dynamical processes on networks.
For example, shortest cycles provide the fastest 
feedback paths in the network
and introduce correlations between the signals arriving
at a given node via different links.
It was found that in neural circuits the lengths of the shortest 
cycles determine the frequencies 
of broadband spontaneous macroscopic neural oscillations
\cite{Vladimirov2012,Goldental2015,Goldental2017}.
In a broader context, feedback processes are affected by the
entire spectrum of cycle lengths, up to the longest possible
length of the Hamiltonian cycles.
The number of cycles of a given length was studied extensively
in Refs. 
\cite{Marinari2004,Bianconi2005,Marinari2006a,Marinari2006b,Klemm2006,Noh2008}.

In the context of network control theory, it was shown that
dynamical processes on complex networks can be identified and
controlled by a small set of 'determining nodes', which can be
identified from the network structure alone, regardless of the
specific properties of the dynamical process.
Moreover, this set must include at least one node from each
one of the feedback loops in the network
\cite{Fiedler2013,Mochizuki2013}.
This approach was recently applied
\cite{Zanudo2017}.
to the analysis of real biological,
technological and social networks, providing predictions for the set
of nodes whose control can push the network dynamics towards any
desired asymptotic state (fixed point, cycle or limit cycle).

Analytical techniques for treating spin models on networks
are mostly exact on tree structures. Utilizing the local tree
structure of random networks, they provide accurate results
for short range properties. However, in order to obtain insight
about collective and long range correlations, one needs to 
take into account the large scale structure, which notably
involves the statistics of loops
as done recently in Refs.
\cite{Montanari2005,Altieri2017}.

\section{Summary}

We presented an analytical approach for the calculation 
of the distribution of shortest 
cycle lengths in configuration model networks.
This approach is based on a fundamental relation between the
distribution of shortest cycle lengths and the distribution 
of shortest path lengths
in such networks.
It employs an analytical approach for the calculation of the distribution
of shortest path lengths, presented in Ref. 
\cite{Nitzan2016}.
We use this approach for the calculation of the DSCL in
Erd{\H o}s-R\'enyi networks, random regular graphs and
scale-free configuration model networks,
and obtain very good agreement with the results of computer simulations.
The mean and standard deviation of the DSCL in these 
networks are also calculated. 
We also obtain a closed form expression for  
the fraction
of nodes which do not reside on any cycle.
While in this paper we have focused on the case of undirected networks,
cycles are known to be important also in directed networks,
in contexts such as gene regulation networks, neural networks 
and food webs 
\cite{Johnson2017}.
It would thus be interesting to study the DSCL on directed networks.
Another interesting direction is the study of properties of long cycles
\cite{Marinari2007}.
In this context, an open question is the distribution
of longest cycle lengths on random networks.

\appendix

\section{The conditional DSPL for short distances}

The conditional DSPLs presented in 
Eqs. (\ref{eq:P})-(\ref{eq:whP}) 
apply for the case in which $\ell > 2$. 
Here we provide 
the expressions for the conditional DSPLs for the
special cases of $\ell=1$ and $2$.
Starting from $\ell=1$, the probability that
two random nodes of degrees $k$ and $k'$ are not connected to
each other is given by

\begin{equation}
P_{\rm PL}(L>1 | k,k') = 1 - \frac{k k'}{(N-1) \langle K \rangle}.
\end{equation}

\noindent
Similarly, when the node of degree $k$ is selected as a random
neighbor of a random node, while the node of degree $k'$ is
a random node, one obtains

\begin{equation}
\widetilde P_{\rm PL}(L>1 | k,k') = 1 - \frac{(k-1) k'}{(N-2) \langle K \rangle}.
\end{equation}

\noindent
Finally, in the case in which both nodes are selected as random neighbors
of random nodes, one obtains

\begin{equation}
\widehat P_{\rm PL}(L>1 | k,k') = 1 - \frac{(k-1) (k'-1)}{(N-3) \langle K \rangle}.
\end{equation}

\noindent
Proceeding to $\ell=2$, one first evaluates the conditional probability 
$P_{\rm PL}(L>2 | L>1;k,k')$, which is given by

\begin{equation}
P_{\rm PL}(L>2 | L>1; k,k') =
\left[ 1 - \sum_{k''=0}^{\infty}
P(k'') \frac{k k' k''(k''-1)}{N(N-1) \langle K \rangle^2}
\right]^{N-2}.
\end{equation}

\noindent
Carrying out the summation and multiplying by
$P_{\rm PL}(L>1| k,k')$, we obtain

\begin{equation}
P_{\rm PL}(L>2 | k, k') = 
1 - \frac{\langle K^2 \rangle k k'}{N \langle K \rangle^2}
+ \mathcal{O} \left( \frac{1}{N^2} \right),
\end{equation}

\noindent
which is valid under the assumption that $\langle K^2 \rangle$
is finite.
Using similar considerations, one can show that

\begin{equation}
\widetilde P_{\rm PL}(L>2 | k, k') = 
1 - \frac{\langle K^2 \rangle (k-1) k'}{N \langle K \rangle^2}
+ \mathcal{O} \left( \frac{1}{N^2} \right)
\end{equation}

\noindent
and

\begin{equation}
\widehat P_{\rm PL}(L>2 | k, k') = 
1 - \frac{\langle K^2 \rangle (k-1) (k'-1)}{N \langle K \rangle^2}
+ \mathcal{O} \left( \frac{1}{N^2} \right).
\end{equation}

\section{The giant cluster in Erd{\H o}s-R\'enyi networks}

In the asymptotic limit 
the ER
network exhibits a percolation transition at $c=1$, such that for
$c<1$
the network consists only of finite components 
while for 
$c>1$
there is a giant cluster. 
At a higher value of the connectivity, namely at 
$c = \ln N$, 
there is a second transition, above which
the giant cluster encompasses 
the entire network and there are 
no non-giant components. 
We denote the probability that a randomly selected node belongs to the
giant cluster by $g=g(c)$.
Clearly, $g(c) = 0$ for $c \le 1$ and $g(c)=1$ for $c > \ln N$.
For intermediate values of $c$, in the range of
$1 < c < \ln N$, 
the probability that a random node belongs
to the giant cluster
is given by 
$1-g =  \exp(-cg)$
\cite{Bollobas2001}.
Solving for $g$, one obtains

\begin{equation}
g(c)= 1 + \frac{W(-c e^{-c})}{c},
\label{eq:g(c)}
\end{equation}

\noindent
where $W(x)$ is the Lambert $W$ function
\cite{Olver2010}.
ER networks exhibit a special property, 
resulting from the Poisson degree distribution,
Eq. (\ref{eq:poisson}), 
which satisfies
$\tilde P(k) = P(k-1)$,
where 
${\widetilde P}(k)$ 
is given by 
Eq. (\ref{eq:tilde}).
This implies that for the Poisson distribution, the two generating functions defined 
in the main text
are identical, namely 
$G_0(x)=G_1(x)=e^{-c(1-x)}$.
As a consequence of 
Eqs. (\ref{eq:tg}) and (\ref{eq:g}), in ER networks
${\tilde g} = g$. 
This means that in ER networks there is no distinction between 
the statistical properties of a random node and a random neighbor
of a random node.

\section{The probability $P(i \in {\rm cycle} | k)$ in Erd{\H o}s-R\'enyi 
networks for $k=4$ and $5$}

Here we present analytical results for 
$P(i \in {\rm cycle}) | k)$
for $k=4$ and $5$ in an ER network, obtained
from Eq. (\ref{eq:cycle_k}).
It is found that

\begin{eqnarray}
P(i \in {\rm cycle}&| K=4)
=
6g^2 - 3g^4 
-12 g^2 \langle g_k^2 \rangle
+12 g^2 \langle g_k^2 \rangle^2
+4 g^3 \langle g_k^3 \rangle
\nonumber \\
&
+4 \langle g_k^2 \rangle^3
-12 g \langle g_k^2 \rangle^2 \langle g_k^3 \rangle
-3 \langle g_k^2 \rangle^4
+6  \langle g_k^2 \rangle^2 \langle g_k^3 \rangle^2
- \langle g_k^3 \rangle^4,
\label{eq:icyc4}
\end{eqnarray}

\noindent
and

\begin{eqnarray}
P(i&\in&{\rm cycle} | K=5)
=
30 g^4 \left\langle g_k^2\right\rangle -5 g^4 
\left\langle g_k^4\right\rangle -15 g^4-60 g^3   
\left\langle g_k^3\right\rangle  \left\langle g_k^2\right\rangle  
\nonumber \\
&+&20 g^3 \left\langle   g_k^3\right\rangle
-70 g^2 \left\langle g_k^2\right\rangle^3+60 g^2 \left\langle   
g_k^2\right\rangle^2+30 g^2 \left\langle g_k^4\right\rangle  
\left\langle g_k^2\right\rangle^2 
-30 g^2 \left\langle g_k^2\right\rangle
\nonumber \\
&+&60 g^2 \left\langle g_k^3\right\rangle^2 
\left\langle g_k^2\right\rangle +10 g^2+12 
\left\langle g_k^2\right\rangle^5-15 
\left\langle g_k^4\right\rangle  \left\langle g_k^2\right\rangle^4
-15 \left\langle g_k^2\right\rangle^4
\nonumber \\
&-&70 \left\langle g_k^3\right\rangle^2 
\left\langle g_k^2\right\rangle^3+10 
\left\langle g_k^4\right\rangle^2 
\left\langle g_k^2\right\rangle^3
+120 g \left\langle g_k^3\right\rangle  
\left\langle g_k^2\right\rangle^3
+10  \left\langle g_k^2\right\rangle^3
\nonumber \\
&+&30 \left\langle g_k^3\right\rangle^2 
\left\langle g_k^2\right\rangle^2
-60 g \left\langle g_k^3\right\rangle  
\left\langle g_k^2\right\rangle^2+60 
\left\langle g_k^3\right\rangle^2 
\left\langle g_k^4\right\rangle  
\left\langle g_k^2\right\rangle^2
\nonumber \\
&-&60 g \left\langle g_k^3\right\rangle  
\left\langle g_k^4\right\rangle  
\left\langle g_k^2\right\rangle^2
+30 \left\langle g_k^3\right\rangle^4 
\left\langle g_k^2\right\rangle 
-60 g\left\langle g_k^3\right\rangle^3 
\left\langle g_k^2\right\rangle  
\nonumber \\
&-&30 \left\langle  g_k^3\right\rangle^2 
\left\langle g_k^4\right\rangle^2 
\left\langle g_k^2\right\rangle
-\left\langle g_k^4\right\rangle^5-5 
\left\langle g_k^3\right\rangle^4+10 
\left\langle g_k^3\right\rangle^2 
\left\langle g_k^4\right\rangle^3
\nonumber \\
&-&15 \left\langle g_k^3\right\rangle^4 
\left\langle g_k^4\right\rangle +20 g 
\left\langle g_k^3\right\rangle^3 
\left\langle g_k^4\right\rangle.
\label{eq:icyc5}
\end{eqnarray}

\noindent
These conditional probabilities can be evaluated 
explicitly,
as a function of $c$,
by inserting the moments

\begin{equation}
\langle g_k^3 \rangle =
1 - 3 e^{-cg} + 3 e^{-cg(2-g)}
- e^{-cg(3-3g+g^2)}
\end{equation}

\noindent
and

\begin{equation}
\langle g_k^4 \rangle =
1 - 4 e^{-cg} + 6 e^{-cg(2-g)}
-4 e^{-cg(3-3g+g^2)}
+e^{-c[1-(1-g)^4]}.
\end{equation}

\noindent
Higher order moments can be obtained from Eq. (\ref{eq:g_k^n}).

\section{The giant cluster in random regular graphs}

In this appendix we show that in random regular 
networks the giant cluster encompasses the entire network,
namely $g=1$.
In the random regular graph, the degree distribution is 
$P(k) = \delta_{k,c}$,
where $\delta_{k,c}$ is the Kronecker symbol and $c$ is an integer.
We will focus on the case of $c \ge 3$.
In this case 

\begin{equation}
G_1(x)=x^{c-1}
\label{eq:g1rrg}
\end{equation}

\noindent
and 

\begin{equation}
G_0(x)=x^c.
\label{eq:g0rrg}
\end{equation}

\noindent
For $c=1$ the random regular graph 
consists of dimers, while for $c=2$
it consists of loops of various lengths.
A fully developed network is obtained only for $c \ge 3$,
and this will be the case of interest in the present work.
To obtain the size of the giant cluster we look for solutions 
of Eq. (\ref{eq:tg}). 
Inserting $G_1(x)$ from Eq. (\ref{eq:g1rrg}) 
into Eq. (\ref{eq:tg}) we obtain 
$1 - \tilde g = (1-\tilde g)^{c-1}$.
It is easy to see that $\tilde g=0$
and $\tilde g = 1$ are solutions of this equation.
Moreover, for $c \ge 3$ the expression on the right hand side is
smaller than the expression on the left hand side for any
$0 < \tilde g < 1$.
Thus, for $c \ge 3$ there are
no other solutions for Eq. (\ref{eq:tg}).
This proves that $\tilde g$ may be either $0$ or $1$.
Inserting these solutions into Eq. (\ref{eq:g}) we find that
in both cases
$g$ is equal to $\tilde g$, namely
$g=0$ or $1$.
The solution $g=0$ stands for the case in which there is no giant 
cluster, while the solution $g=1$ implies that the giant cluster
encompasses the entire network.
In order to determine which of these possible solutions is the relevant one,
we use the criterion of Molloy and Reed for the existence of a giant 
cluster
\cite{Molloy1995,Molloy1998}. 
It states that if
$\langle K^2 \rangle > 2 \langle K \rangle$
then there is a giant cluster, namely $g > 0$.
In the case of a random regular graph 
$\langle K \rangle = c$
and 
$\langle K^2 \rangle = c^2$.
Thus, for $c \ge 3$ the 
Molloy and Reed
criterion is satisfied and $g > 0$.
Hence, the only possible solution is $g=1$, namely
the giant cluster of a random regular graph with $c \ge 3$ 
encompasses the entire network.

\section{The giant cluster in scale-free networks}

In this appendix we consider a configuration model 
with a power-law degree
distribution of the form
$P(k) \sim k^{-\gamma}$,
where the degrees are bounded in the range
$k_{\rm min} \le k \le k_{\rm max}$.
The normalized degree distribution is given by
Eq. (\ref{eq:PLnorm}).
The mean degree is

\begin{equation}
\langle K \rangle = 
\frac{ \zeta(\gamma-1,k_{\rm min}) - \zeta(\gamma-1,k_{\rm max}+1) }
{ \zeta(\gamma,k_{\rm min}) - \zeta(\gamma,k_{\rm max}+1) },
\label{eq:Kmsf}
\end{equation}

\noindent
while the second moment of the degree distribution is

\begin{equation}
\langle K^2 \rangle = 
\frac{ \zeta(\gamma-2,k_{\rm min}) - \zeta(\gamma-2,k_{\rm max}+1) }
{ \zeta(\gamma,k_{\rm min}) - \zeta(\gamma,k_{\rm max}+1) }.
\label{eq:K2msf}
\end{equation}

\noindent
For $\gamma \le 2$ the mean degree diverges when 
$k_{\rm max} \rightarrow \infty$.
For $2 < \gamma \le 3$ the mean degree is 
bounded while the second moment,
$\langle K^2 \rangle$, diverges.
For $\gamma > 3$ both moments are bounded.
It can be shown that for $\gamma > 2$ and
$k_{\rm min} \ge 2$ 
(where nodes of degrees $0$ and $1$ do not exist),
$\langle K^2 \rangle > 2 \langle K \rangle$
namely the Molloy and Reed criterion is satisfied
and the network exhibits a giant cluster
\cite{Molloy1995,Molloy1998}. 
Below we show that under these conditions
the giant cluster encompasses the entire network.
Inserting a power-law degree distribution
with $k_{\rm max} \rightarrow \infty$
into Eq. 
(\ref{eq:tg})
we obtain

\begin{equation}
G_1(x) = \frac{\Phi(x,\gamma-1,k_{\rm min})}{\zeta(\gamma-1,k_{\rm min})}
x^{k_{\rm min}-1},
\end{equation}

\noindent
where $\Phi(x,\gamma,k)$
is the Lerch transcendent
\cite{Gradshteyn2000}.
It can be shown that for any $0 < x < 1$ the relation
$\Phi(x,s,k) < \Phi(1,s,k) = \zeta(s,k)$ 
is satisfied,
provided that $k >0$.
Therefore,
$G_1(x) < x^{k_{\rm min}-1}$ 
for any value of $x$ in the range
$0 < x <1$.
In the case in which $k_{\rm min} \ge 2$ the inequality
$x^{k_{\rm min}-1} < x$ is also satisfied for $0 < x < 1$.
Thus, $x=0$ and $x=1$ are the only fixed points 
of the generating function $G_1(x)$.
Inserting $1-\tilde g$ instead of $x$ 
and using the criterion of Molloy and Reed
\cite{Molloy1995,Molloy1998},
we find that the only possible value of
$\tilde g$ is $\tilde g = 1$.
Inserting this value in Eq.
(\ref{eq:g})
one finds that $g=1$,
namely that in scale free configuration model networks
with $\gamma > 2$ and $k_{\rm min} \ge 2$, 
the giant cluster encompasses the entire network.

\clearpage
\newpage

\begin{table}
\caption{The different tail distributions for the DSPL and DSCL and the
equations from which each one of them can be evaluated}
%\begin{center}
\begin{tabular}{| l  | l | l | }
\hline \hline
{\bf Distribution}  & {\bf Equation} & {\bf Description}    \\ 
\hline \hline 
$P_{\rm PL}(L>\ell | L>\ell-1)$   &  {\bf Eq. (\ref{eq:P_rec2}) }   & Condidtional DSPL between 
pairs of random nodes$^{\ast}$   \\ \hline
$\widetilde P_{\rm PL}(L>\ell | L>\ell-1)$   &  {\bf Eq. (\ref{eq:P_rec2s}) } &  Conditional DSPL
between random nodes and a RNRNs$^{\dag,}$$^{\ast}$     \\ \hline
$\widehat P_{\rm PL}(L>\ell | L>\ell-1)$   &  {\bf Eq. (\ref{eq:P_rec2sh}) }  & Conditional DSPL
between pairs of RNRNs$^{\ast}$     \\ \hline
\hline
$P_{\rm PL}(L>\ell )$   &  {\bf Eq. (\ref{eq:ProdCond}) }   & DSPL between pairs random nodes$^{\ast}$    \\ \hline
$\widetilde P_{\rm PL}(L>\ell )$   &  {\bf Eq. (\ref{eq:prodtilde}) }   & DSPL between
random nodes and RNRNs$^{\ast}$    \\ \hline
$\widehat P_{\rm PL}(L>\ell )$   &  {\bf Eq. (\ref{eq:prodh}) }   & DSPL between pairs of RNRNs$^{\ast}$     \\ \hline
\hline
$Q_{\rm PL}(L>\ell )$   &  {\bf Eq. (\ref{eq:Q}) }  & Overall DSPL between pairs of random nodes      \\ \hline
$\widetilde Q_{\rm PL}(L>\ell )$   &  {\bf Eq. (\ref{eq:Qt}) }  & Overall DSPL between random
nodes and RNRNs      \\ \hline
$\widehat Q_{\rm PL}(L>\ell )$   &  {\bf Eq. (\ref{eq:Qw}) }   & Overall DSPL between pairs of
RNRNs    \\ \hline
\hline
$P_{\rm PL}(L>\ell | k,k')$   &  {\bf Eq. (\ref{eq:P}) }  & DSPL between pairs of random nodes
of degrees $k$ and $k'$ $^{\ast}$      \\ \hline
$\widetilde P_{\rm PL}(L>\ell  | k,k')$   &  {\bf Eq. (\ref{eq:wtP}) }    & DSPL between random
nodes and RNRNs of degrees $k$ and $k'$ $^{\ast}$   \\ \hline
$\widehat P_{\rm PL}(L>\ell | k,k' )$   &  {\bf Eq. (\ref{eq:whP}) } & DSPL between pairs of RNRNs of degrees $k$ and $k'$ $^{\ast}$        \\ \hline
\hline
$Q_{\rm PL}(L>\ell  | k,k')$   &  {\bf Eq. (\ref{eq:Q2}) }  & Overall DSPL between pairs of random
nodes of degrees $k$ and $k'$      \\ \hline
$\widetilde Q_{\rm PL}(L>\ell | k,k' )$   &  {\bf Eq. (\ref{eq:Qtc}) }   & Overall DSPL between
random nodes and RNRNs of degrees $k$ and $k'$  \\ \hline
$\widehat Q_{\rm PL}(L>\ell  | k,k')$   &  {\bf Eq. (\ref{eq:Qhc}) }    & Overall DSPL between
pairs of RNRNs of degrees $k$ and $k'$    \\ \hline
\hline
$P_{\rm CL}(L>\ell | k)$   &  {\bf Eq. (\ref{eq:dscl_c1}) }   & DSCL of nodes of degree $k$    \\ \hline
$P_{\rm CL}(L>\ell )$   &  {\bf Eq. (\ref{eq:dscl})} &  DSCL    \\ \hline
\hline
\end{tabular}
\label{table}
%\end{center}
\end{table}
\noindent
$^{\ast}$ For pairs of nodes which reside on the same connected component; 
\\
\noindent
$^{\dag}$ RNRNs: random neighbors of random nodes;

\clearpage
\newpage

\begin{figure}
\includegraphics[width=8cm]{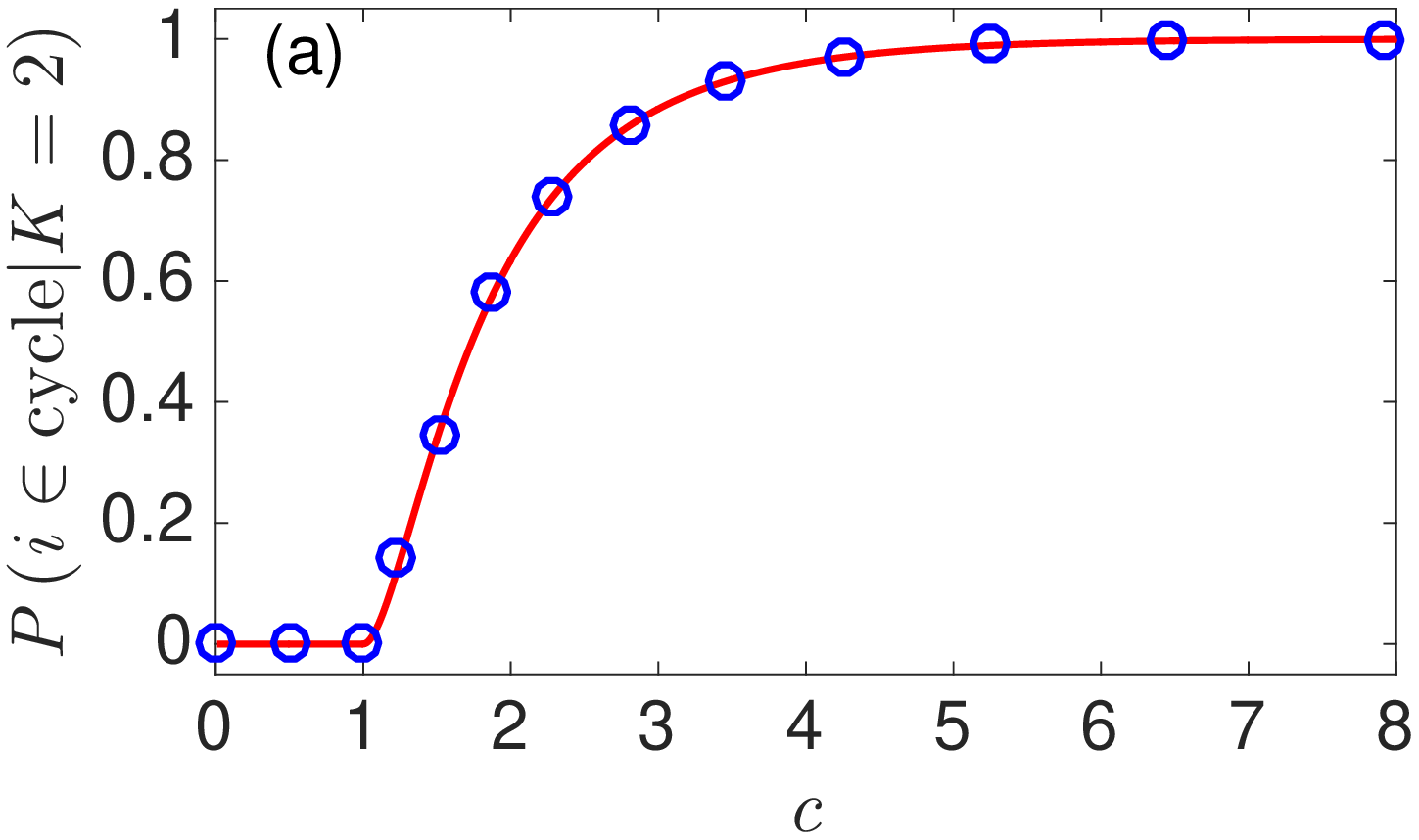} \\
\includegraphics[width=8cm]{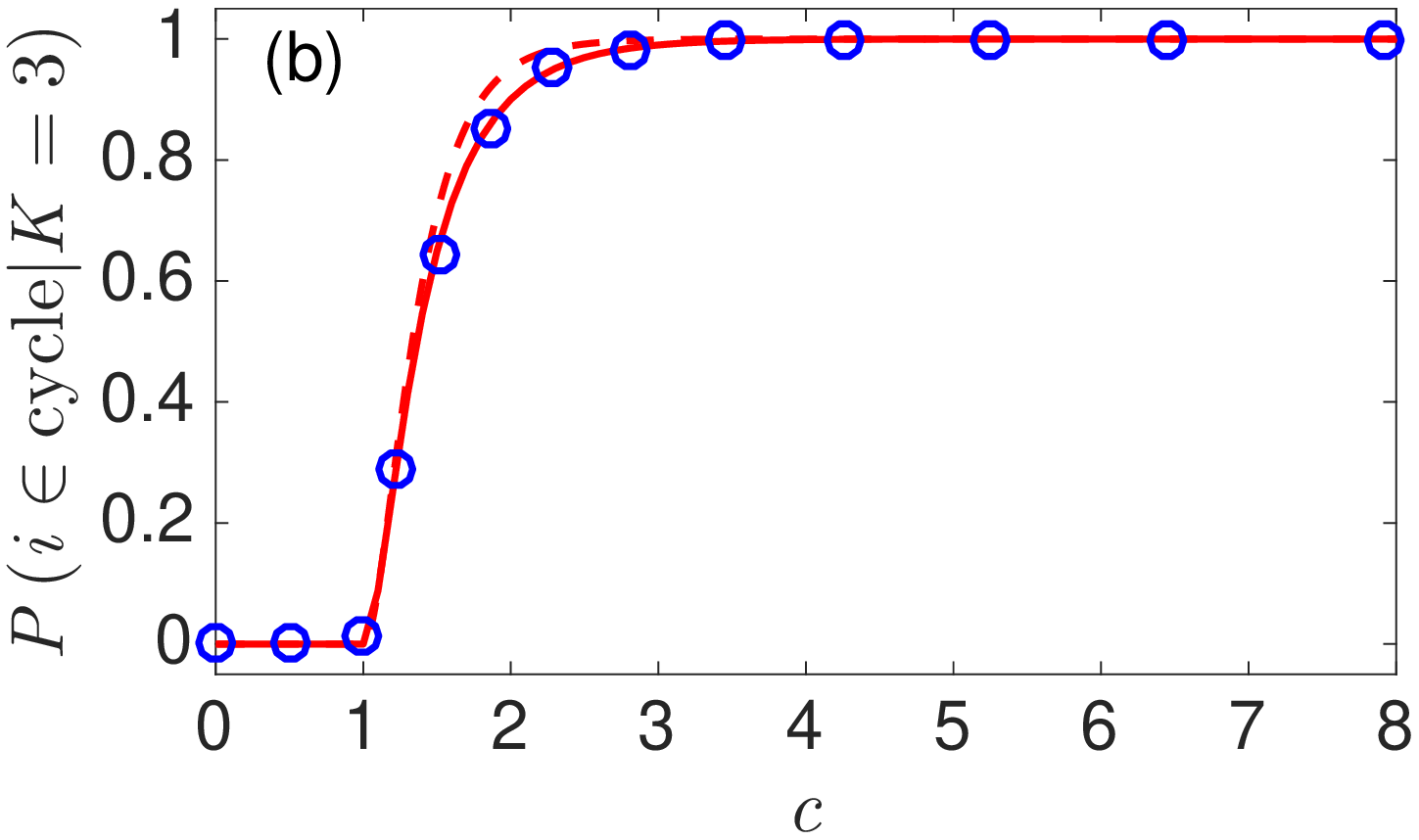} \\
\includegraphics[width=8cm]{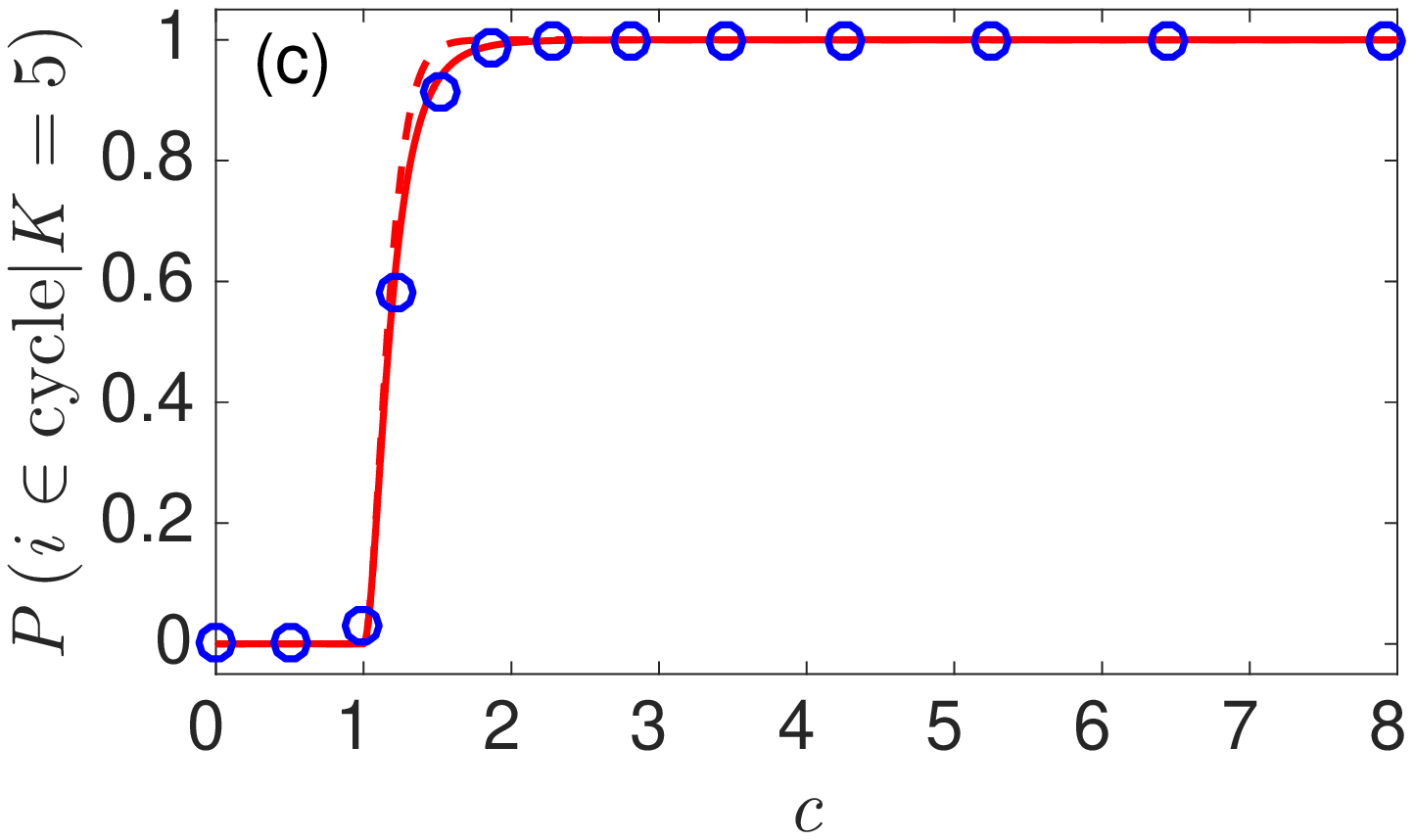}
\caption{
(Color online)
The conditional probability $P(i \in {\rm cycle} | K=k)$ 
that a random node, $i$, 
of degree $k=2$ (a), $k=3$ (b) and $k=5$ (c) 
resides on at least one cycle, 
as a function of the mean degree, $c$,
in an ER network of $N=10^4$ nodes.
For $0 < c < 1$ there are no cycles and therefore
$P(i \in {\rm cycle} | K=k)=0$ .
For $c > 1$ the probability that a random node, $i$, of
a given degree, $k$, resides on at least one cycle increases monotonically with $c$.
This is due to the fact that as $c$ is increased the degrees of its neighbors increase
and they are thus more likely to be connected to each other on the reduced network
which does not include the node $i$.
The analytical results 
obtained from the
simpler approach, 
described by
Eq. (\ref{eq:cyck}), 
are shown in dashed lines,
while the analytical results obtained from 
the more detailed approach, 
described by
Eq. (\ref{eq:Pncc2}), 
are shown in solid lines.
Incidentally, the two analytical curves coincide for $k=2$,
while for $k=3$ and $5$, the more detailed theory is found to
be in better agreement with the
results of computer simulations (circles).
}
\label{fig:5}
\end{figure}

\end{document}